\documentclass[11pt]{article}

\usepackage[T1]{fontenc}
\usepackage[utf8]{inputenc}
\usepackage{lmodern}
\usepackage[margin=1in]{geometry}
\usepackage{amsmath,amssymb}
\usepackage{bm}
\usepackage{graphicx}
\usepackage{booktabs}
\usepackage{subcaption}
\usepackage[round,authoryear]{natbib}
\usepackage{authblk}
\usepackage{xcolor}
\usepackage{tikz}
\usetikzlibrary{calc}
\usepackage[colorlinks=true,linkcolor=blue!60!black,citecolor=blue!60!black,urlcolor=blue!60!black]{hyperref}

\newcommand\doverline[1]{%
\tikz[baseline=(nodeAnchor.base)]{
    \node[inner sep=0] (nodeAnchor) {$#1$};
    \draw[line width=0.1ex,line cap=round]
        ($(nodeAnchor.north west)+(0.0em,0.2ex)$)
            --
        ($(nodeAnchor.north east)+(0.0em,0.2ex)$)
        ($(nodeAnchor.north west)+(0.0em,0.5ex)$)
            --
        ($(nodeAnchor.north east)+(0.0em,0.5ex)$)
    ;
}}

\newcommand{\dwidetilde}[1]{%
  \begingroup
  \setbox0=\hbox{$\widetilde{#1}$}%
  \rlap{%
    \smash{%
      \raise0.35ex\hbox to \wd0{%
        \hss$\widetilde{\phantom{#1}}$\hss
      }%
    }%
  }%
  \box0%
  \endgroup
}

\newcommand{\solidline}{\tikz[thick, baseline=-0.5ex]{\draw[black] (0,0) -- (0.6,0);}}
\newcommand{\dashedline}{\tikz[thick, baseline=-0.5ex]{\draw[black, dashed] (0,0) -- (0.6,0);}}
\newcommand{\dashdotline}{\tikz[thick, baseline=-0.5ex]{\draw[black, dash dot] (0,0) -- (0.6,0);}}

\title{Aggregation, breakup, and size-dependent transport in a turbulent channel flow with cohesive particles}

\author[1]{Alexandre D. Leonelli\thanks{Corresponding author: \href{mailto:adleonelli@ucsb.edu}{adleonelli@ucsb.edu}}}
\author[1,2]{Lukas Widmer}
\author[1]{Eckart Meiburg}
\affil[1]{Department of Mechanical Engineering, University of California Santa Barbara, Santa Barbara, CA, USA}
\affil[2]{Department of Mechanical and Process Engineering, ETH Zurich, Zurich, Switzerland}
\date{}

\begin{document}
\maketitle

\begin{abstract}
Due to attractive inter-particle forces, cohesive particles suspended in turbulence undergo a complex process of aggregation, breakup, and restructuring. Despite a growing body of knowledge on the ``flocculation'' of cohesive granular materials suspended in homogeneous isotropic turbulence, little focus has so far been placed on wall-bounded flows where turbulence and shear are inhomogeneous. This study presents a first investigation of a fully developed wall-bounded flow of resolved cohesive particles. Five direct numerical simulations of turbulent channel flows laden with finite-sized particles at successively increasing cohesive strength are performed. A population balance equation (PBE) framework is used to analyze aggregate dynamics. When integrated over the full domain, the PBE is closed by aggregation and breakup alone. However, this balance is found to not hold locally in the wall-normal direction, where regions of net aggregate production and depletion are identified. This imbalance is shown to be compensated by the size-dependent wall-normal transport of aggregates, revealing a mean circulation: larger aggregates are preferentially produced in the channel center and migrate toward the wall where they break, while smaller aggregates are transported away from the wall, grow, and reenter the cycle.
\end{abstract}



\section{Introduction}
Suspensions of cohesive materials are ubiquitous in many engineered and environmental systems. The complex process of aggregation, breakup, and restructuring, known in sediment transport as ``flocculation,'' is poorly understood and difficult to model \citep{winterwerp_book_2004}. A popular approach, inspired by \cite{smoluchowski_1916,smoluchowski_1918}, is the population balance equation (PBE). Also used to model bubble and droplet coalescence and breakup, and reacting flows \citep[e.g.,][]{shaw_droplets_PBE_2003, rigopoulos_pbe_reacting_2007,ramkrishna_pbe_2014}, the PBE describes the evolution of aggregate size concentration over time and space by an advection equation with source and sink terms accounting for discrete aggregation and breakup events. These terms involve probabilistic and rate-describing kernels which are parameterized to construct low-order models \citep{ives_pbe_1973, winterwerp_flocculation_1998}. Despite growing interest in incorporating boundary layer effects \citep{penaloza-giraldo_bbl_2025, penaloza-giraldo_bbl_2025-2}, parameterizations are often over-simplified, or agnostic to local flow conditions.

Experimental studies have provided valuable insights into aggregate morphology and population dynamics; however, direct observation of aggregation and breakup events requires high-speed imaging and well-controlled environments, often making such experiments prohibitive. Notably, \citet{saha_breakup_2016} captured Lagrangian breakup statistics in quasi-HIT, while \citet{hoffman_experimental_2023} reported statistics on individual aggregation and breakup events in a humid turbulent flow laden with glass particles. Still, most experimental investigations are limited to single aggregates or simplified flows \citep{blaser_flocs_2000, dizaji_shear_2019, qi_breakup_2025}, and systematic variation of cohesive forces remains challenging \citep{gans_ccgm_2020}. These difficulties have motivated a growing body of numerical work.

With access to detailed data, particle-resolved simulations have yielded insight into how turbulence intensity, cohesive strength, particle volume fraction, and other micromechanical properties influence flocculation in HIT \citep[e.g.][]{zhao_flocculation_2021, yao_deagglomeration_2021, manning_flocculation_2024} and, to a lesser extent, wall-bounded flows \citep[e.g.][]{afkhami_fully_2015, njobuenwu_large_2018, schutte_formation_2018}. A key contribution in this direction is, \citet{babler_numerical_2015}, who conducted one-way coupled DNS of point-like aggregates in various wall-bounded flows and identified two breakup mechanisms: weaker aggregates broke throughout the flow due to local turbulent stresses, whereas stronger aggregates migrated closer to a wall before breaking under stresses induced by mean shear. With a similar numerical approach, \citet{breuer_modeling_2015} investigated a turbulent channel laden with aggregating particles (no breakup) and found that aggregation rates peak near the wall, where collision rates are highest. These findings have been extended by studies that incorporate both aggregation and breakup, however, due to computational limitations at the time, those studies were limited to initial aggregation-dominated transients \citep{afkhami_fully_2015, njobuenwu_large_2018, haervig_early_2018}.

To avoid complications from drag modeling \citep{manning_flocculation_2024}, the present study employs particle-resolved direct numerical simulations (PR-DNS) to investigate the aggregation and breakup of cohesive particles suspended in a turbulent channel flow. Despite independent insight into each process, to the best of the authors' knowledge, no studies demonstrate if aggregation and breakup balanced locally in the wall-normal direction, which would reveal regions of net aggregate production or depletion. We employ a PBE framework, introduced in section \ref{sec:pbe}, to process the simulation results and develop a statistical description of aggregation and breakup in the wall-normal direction. Following bulk considerations in section \ref{sec:bulk}, regions of net aggregate production and depletion are identified in both physical and size-space in section \ref{sec:wall-normal-balance}, revealing a wall-normal imbalance between aggregation and breakup. We then show that this imbalance is closed by the wall-normal advection of aggregates, giving rise to a conceptual picture of statistical circulation in the joint space of size and wall-normal coordinate. Finally, to clarify the mechanisms driving this circulation, near-wall aggregate breakup is examined in section \ref{sec:agg-fate} while aggregate orientation, rotation, and deformation are quantified in section \ref{sec:axial_strain_rate}.

\section{Numerical approach}\label{sec:numerical-approach}
The fluid phase is governed by the continuity
\begin{equation}\label{eqn:cont}
  \boldsymbol{\nabla}\cdot\boldsymbol{u} = 0,
\end{equation}
and Navier Stokes equations,
\begin{equation}\label{eqn:nse}
  \frac{\partial \boldsymbol{u}}{\partial t} + \boldsymbol{\nabla}\cdot\left(\boldsymbol{u}\boldsymbol{u}\right) = -\frac{1}{\rho_{f}}\boldsymbol{\nabla} p + \nu \nabla^2\boldsymbol{u} + \boldsymbol{f}_{IBM},
\end{equation}
for an incompressible Newtonian fluid with density $\rho_f$ and kinematic viscosity $\nu$. The fluid velocity at a point $\boldsymbol{x} = (x, y, z)$ and time $t$ is denoted by $\boldsymbol{u}(\boldsymbol{x}, t) = (u, v, w)$, while $p$ denotes the pressure, and $\boldsymbol{f}_{IBM}$ the numerical volume force responsible for the coupling between fluid and particle by means of the immersed boundary method (IBM). The particle linear,
\begin{equation}\label{eqn:particle-linear-momentum}
    m_p \frac{\text{d}\boldsymbol{u}_i^P}{\text{d}t} = \oint_{\Gamma_i}\boldsymbol{\tau}\cdot\boldsymbol{n}_i^P\text{d}A + \boldsymbol{F}_{c,i},
\end{equation}
and angular momentum equations,
\begin{equation}\label{eqn:particle-angular-momentum}
    I_p\frac{\text{d}\boldsymbol{\omega}_i^P}{\text{d}t} = \oint_{\Gamma_i}\boldsymbol{r}_i^P\times\left(\boldsymbol{\tau}\cdot\boldsymbol{n}_i^P\right)\text{d}A + \boldsymbol{T}_{c,i},
\end{equation}
are solved concurrently. Here, $\boldsymbol{u}_i^P(t)=(u_i^P, v_i^P, w_i^P)$ and $\boldsymbol{\omega}_i^P(t)=(\omega_{x,i}^P,\omega_{y,i}^P,\omega_{z,i}^P)$ are the linear and angular velocities of particle $i$, $\Gamma_i$ its surface, and $\boldsymbol{r}_i^P=\boldsymbol{x}-\boldsymbol{x}_i^P$ is the displacement vector from its center, $\boldsymbol{x}_i^P(t)=(x_i^P,y_i^P,z_i^P)$. All particles have mass $m_p$, diameter $d$, density $\rho_p$, and moment of inertia $I_p=\pi\rho_p d^5/60$. The hydrodynamic stress tensor $\boldsymbol{\tau} = p\boldsymbol{\mathsf{I}} + 2\rho_f\nu\boldsymbol{\mathsf{E}}$ with $\boldsymbol{\mathsf{E}} = \left(\nabla \boldsymbol{u} + \nabla \boldsymbol{u}^T\right) / 2$ is integrated over $\Gamma_i$, while the contact forces, $\boldsymbol{F}_{c,i}$, and torques, $\boldsymbol{T}_{c,i}$, act when appropriate.

Importantly included in $\boldsymbol{F}_{c,i}$ are the cohesion and lubrication forces, $\boldsymbol{F}_{coh,ij}$ and $\boldsymbol{F}_{\ell,ij}$. In this work, $\boldsymbol{F}_{coh,ij}$ is modeled using the formulation of \cite{vowinckel_settling_2019},
\begin{equation}\label{eqn:cohesion-model}
\boldsymbol{F}_{coh,ij} =
\begin{cases}
    -\dfrac{A_H d}{2\zeta_0 \lambda^3}\left(\zeta_{ij}^2 - \zeta_{ij}\lambda\right)\boldsymbol{e}_{ij}^P, & 0 < \zeta_{ij} \leq \lambda,\\
0, & \text{otherwise},
\end{cases}
\end{equation}
which provides a computationally efficient approximation of the short-range attractive forces predicted by the Derjaguin--Landau--Verwey--Overbeek (DLVO) theory of van der Waals (vdW) interactions. Here, $\boldsymbol{e}_{ij}^P$ is the unit vector directed from the center of particle $i$ to particle $j$, $\zeta_{ij} = \lVert\boldsymbol{x}_i^P-\boldsymbol{x}_j^P\rVert-d$ is the minimum surface-to-surface distance between the pair, and $\zeta_0$ represents the scale of microscopic surface asperities. The parameter $\lambda$ defines the finite range over which the cohesive force is distributed and must be set comparable to the other length scales in the problem to not impose computational constraints, thus $\lambda = d/20$, as in \citet{vowinckel_settling_2019}, who show that pairwise interactions are relatively insensitive to $\lambda$. The Hamaker constant, $A_H$, characterizes the net short-range adhesion between particles, encapsulating the combined influence of particle and fluid material properties. Interpreted here as an effective value, it may therefore also represent additional adhesive mechanisms beyond vdW interactions alone such as polymer bridging or biofilm-mediated cohesion.

When $\zeta_{ij} \leq 2\Delta x$, the close-range hydrodynamic forces can no longer be resolved in the normal direction. Following \citet{vowinckel_settling_2019} and \citet{biegert_collision_2017}, the unresolved normal lubrication force is modeled as
\begin{equation}\label{eq:lub-model}
    \boldsymbol{F}_{\ell,ij} = \begin{cases}
        -\frac{3\pi \rho_f \nu d^2}{8\max(\zeta_{ij}, \zeta_{min})}\boldsymbol{g}_{ij} & 0 < \zeta_{ij} \leq 2 \Delta x,\\
        0 & \text{otherwise},
    \end{cases}
\end{equation}
where $\Delta x$ is the isotropic grid spacing, $\boldsymbol{g}_{ij} = \left[(\boldsymbol{u}_i^P - \boldsymbol{u}_j^P)\cdot \boldsymbol{e}_{ij}^P\right]\boldsymbol{e}_{ij}^P$ is the relative normal velocity between the two particles, and $\zeta_{min} = 1.5\times10^{-3}d$ is a regularizing constant calibrated by \citet{biegert_collision_2017}.

The cohesion and lubrication forces are accompanied by normal and tangential collision forces, $\boldsymbol{F}_{n,ij}$ and $\boldsymbol{F}_{t,ij}$. Accumulated, the contact forces and torques acting on particle $i$ are
\begin{equation}\label{eq:forces}
\boldsymbol{F}_{c,i} = \boldsymbol{F}_{\ell,iw} + \boldsymbol{F}_{n,iw} + \boldsymbol{F}_{t,iw} + \sum_{j,j\neq i}\left(\boldsymbol{F}_{\ell,ij} + \boldsymbol{F}_{n,ij} + \boldsymbol{F}_{t,ij} + \boldsymbol{F}_{coh,ij}\right),
\end{equation}
and
\begin{equation}\label{eq:torques}
    \boldsymbol{T}_{c,i} = \frac{d}{2}\boldsymbol{e}_{iw}^P\times\boldsymbol{F}_{t,iw} + \sum_{j,j\neq i}\frac{d}{2}\boldsymbol{e}_{ij}^P\times\boldsymbol{F}_{t,ij},
\end{equation}
where the definitions above extend to interactions between particle $i$ and wall $w$ following \citet{biegert_collision_2017}, with the exception of the cohesive force, which acts only between particles, consistent with this work's focus on the dynamics of suspended aggregates rather than their capture by walls. The collision and lubrication models, also adopted in \citet{vowinckel_settling_2019}, were calibrated against experiment by \citet{biegert_collision_2017} so that collision kinematics and a restitution coefficient of $0.97$ are well reproduced.

\section{Physical configuration}\label{sec:config}
\begin{figure}[h!]
    \centering
    \includegraphics[width=0.7\linewidth]{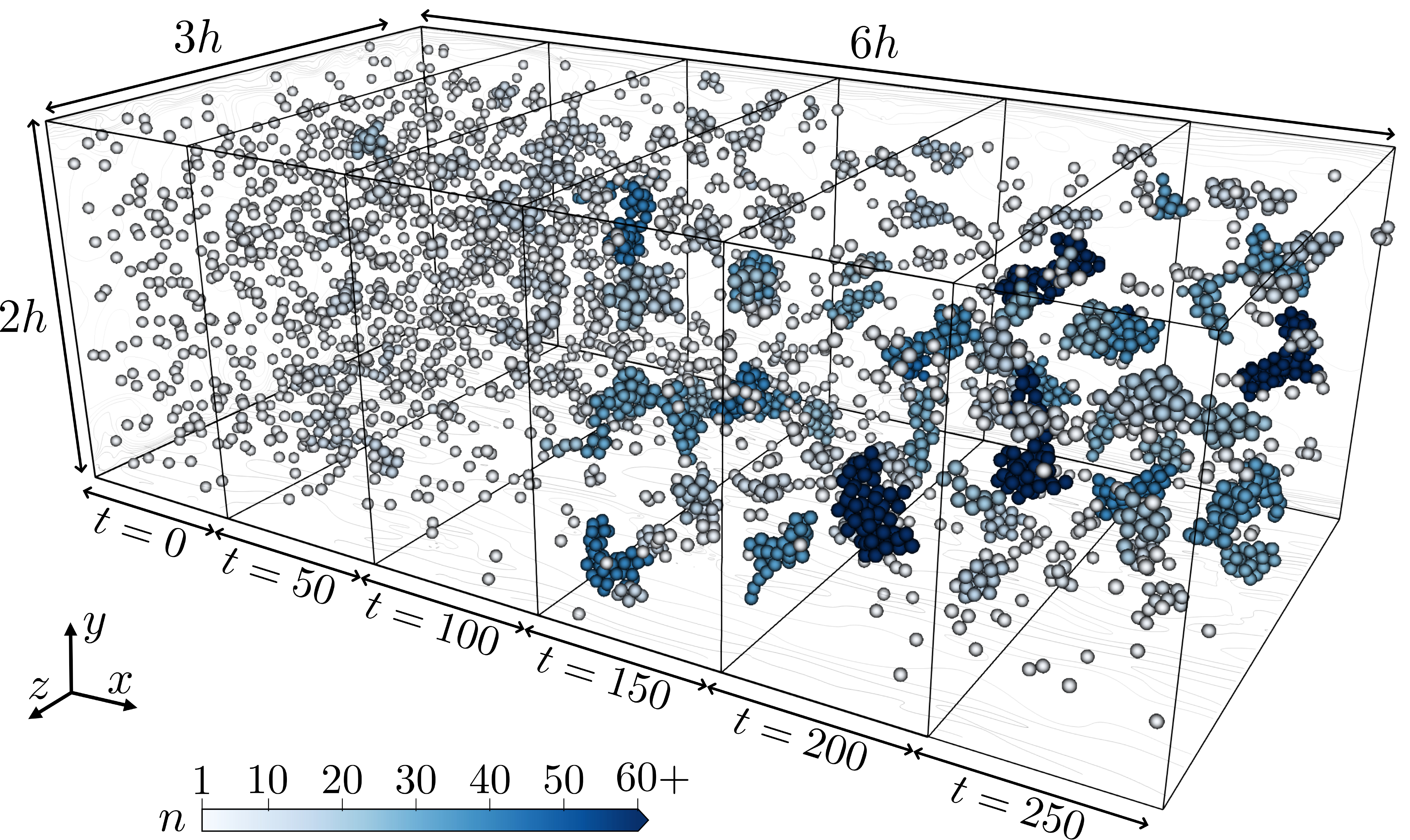}
    \caption{Schematic of the simulation domain. Slices from $Co=1$ are shown sequentially along the channel for increasing times, with aggregates colored by the number of constituent particles. A corresponding video from case $Co=0.48$ can be found in the supplementary materials.}
    \label{fig:schematic}
\end{figure}
A channel of size $(L_x, L_y, L_z) = (6h, 2h, 3h)$, half-height $h$, is filled with a Newtonian fluid laden with $N = 3841$ (volume fraction $\phi = N \pi d^3 / (6 L_x L_y L_z) = 0.015$) cohesive particles of diameter $d=2h/31$ and density ratio $\rho_s = \rho_p / \rho_f = 1$, cf. figure~\ref{fig:schematic}. The streamwise ($x$) and spanwise ($z$) directions are periodic while no-slip walls are imposed at $y=0$ and $y=2h$. An adaptive pressure gradient maintains a constant bulk velocity $U$ and Reynolds number $\mathit{Re}=U h / \nu = 2800$, which, when formulated with the friction velocity $u_\tau = \sqrt{\tau_w / \rho_f}$ and wall shear stress $\tau_w$, corresponds to the well-studied friction Reynolds number of $\mathit{Re}_\tau = u_\tau h / \nu = 180$ of \citet{kim_1987}. The domain is discretized using a cubic mesh with grid spacing $\Delta x = h/186$ so that the characteristic length scale at the wall, the viscous unit $\delta_\nu = \nu / u_\tau$, is sufficiently resolved: $\delta_\nu / \Delta x = 0.9678$. The particle boundary layers are also resolved, even those at the extreme ends of aggregates. To verify this we follow the procedure detailed in appendix \ref{sec:slip-vel} to compute the effective slip velocity between particle $i$ and the fluid, $\Delta\boldsymbol{u}_i^{PF}$, and subsequently the particle Reynolds number, $\mathit{Re}_{p,i} = \lVert \Delta\boldsymbol{u}_i^{PF}\rVert d / \nu$. Over all cases, $\max_i \mathit{Re}_{p,i} \leq 49.58$, so the minimum particle boundary layer is estimated to be $\min_i \delta_{p,i} \approx d / \sqrt{\max_i\mathit{Re}_{p,i}} = 1.65\delta_\nu$ and is therefore resolved. Furthermore, the particle resolution of $d / \Delta x = 12$ matches past IBM based PR-DNS studies in turbulent channel flow at a minimum of $\mathit{Re}_\tau = 180$ \citep[and references therein]{costa_small_particles_2020}. It should also be noted that the ratio of particle diameter to Kolmogorov length scale is large in this work, with $d / \eta \approx 6.78$ at $y = 0.06h$ and $d / \eta \approx 3.15$ at the centerline on average. While $\eta$ is appropriately resolved, the implications of the particle size are discussed in the conclusion. Lastly, validation for a single-phase and cohesionless suspension is discussed in section \ref{sec:validation}, with detailed validation of the hydrodynamics and collision schemes against experimental observations reported in \citet{biegert_collision_2017}, to excellent agreement. Further convergence analysis and comparison to empirical and theoretical settling velocities for single- and multi-sphere aggregates at particle Reynolds numbers up to $\mathit{Re}_p = 100$ can be found in \citet{vowinckel_settling_2019} and \citet{maches_settling_2024}.

The cohesive force, equation \eqref{eqn:cohesion-model}, is characterized by the cohesive number,
\begin{equation}
    Co = \frac{\max\lVert\boldsymbol{F}_{coh,ij}\rVert}{\rho_f U^2d^2},
\end{equation}
which quantifies the ratio of maximum cohesive force to a characteristic particle inertial force. In order to observe measurable aggregation for the present particle size and $\mathit{Re}$ we select $Co = \{0, 0.24, 0.48, 1.0, 1.5\}$. As mentioned in section \ref{sec:numerical-approach}, given the limitations on particle size in PR-DNS, the Hamaker constant should be interpreted as an effective constant; taking the properties of water, $\nu = 1.002\times 10^{-6}\,\text{m}^2/\text{s}$ and $\rho_f = 1000\,\text{kg/m}^3$, and a surface asperity scale of $\zeta_0 = 0.2\,\text{nm}$ \citep{Israelachvili_asperities}, yields $A_H = \mathcal{O}(10^{-16}\text{-}10^{-15})\,\text{J}$, whereas $A_H = \mathcal{O}(10^{-20}\text{-}10^{-18})\,\text{J}$ for materials such as quartz in water \citep{Bergstrom_hamaker, Marshall_Li_2014}.

Particles are initially distributed randomly, without overlap, throughout a fully developed single-phase channel flow, with each particle assigned the mean linear and angular velocities of the replaced fluid. At time $t=0$ the particles are released and the simulations are marched forward, allowing particles to aggregate, fragment, and redistribute in the turbulent flow. The fluid is advanced using a time step $\Delta t_f$ that is continuously adapted so that the Courant number, $\mathcal{C}$, satisfies the Courant--Friedrichs--Lewy condition, $\mathcal{C} \leq 0.5$. Particle kinematics are evaluated using a sub-stepping routine with time step $\Delta t_p = \Delta t_f / 15$. In the spatiotemporal schematic of figure~\ref{fig:schematic}, the domain is partitioned into six temporal segments along the streamwise direction to illustrate aggregate growth.

\section{Validation}\label{sec:validation}

\subsection{Single-phase validation}
\begin{figure}
    \centering
    \includegraphics[width=0.9\linewidth]{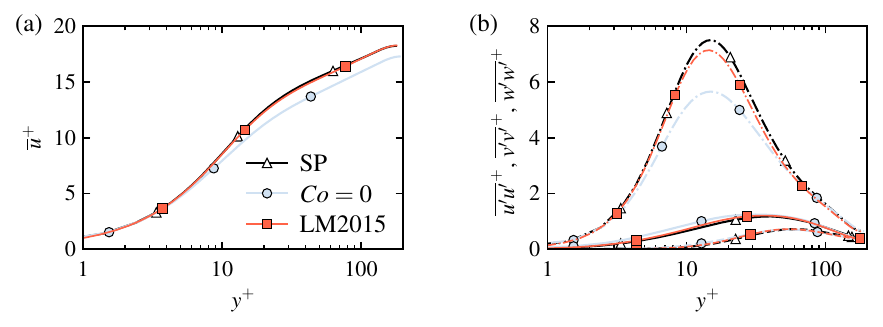}
    \caption{Comparison of (a) the mean streamwise fluid velocity, $\overline{u}^+$ and (b) velocity fluctuations, $\overline{u'u'}^+$; \protect\dashdotline, $\overline{v'v'}^+$; \protect\solidline, and $\overline{w'w'}^+$; \protect\dashedline, from the single-phase (SP) and non-cohesive ($Co = 0$) cases from this study, and the channel flow DNS of $\mathit{Re}_\tau=182.088$ by \citet{lee_moser_turbulence_2015} (LM2015).}
    \label{fig:sp_val}
\end{figure}

The particle-laden flow is initialized from a fully developed single-phase turbulent channel flow with a measured $\mathit{Re}_\tau = 179.82$. For validation, the wall-normal profiles of the mean velocity, $\overline{u}^+$ and velocity fluctuations, $\overline{u'u'}^+$, $\overline{v'v'}^+$, and $\overline{w'w'}^+$, from the single-phase precursor simulation; SP, are displayed in figure \ref{fig:sp_val} along with the DNS at $\mathit{Re}_\tau = 182.088$ of \citet{lee_moser_turbulence_2015} (LM2015). Here, $\overline{\,\cdot\,}$ denotes a plane and time averaging while a superscript ${\cdot}^+$ denotes a quantity made dimensionless with $\delta_\nu$ and $u_\tau$. Small discrepancies are observed near $y^+ = 10$, likely due to the second order accuracy of the present solver as opposed to the seventh-order method employed by \citet{lee_moser_turbulence_2015}. For the purpose of comparison, the domain size is selected to match a history of PR-DNS of turbulent channel flow \citep[cf.][and references therein]{picano_turbulent_2015, mirbod_turbulent_2023}. Though too small to contain the large scale motions observed in extended domains, the present size has been shown to accurately capture one-point statistics in both single-phase \citep{lozano_duran_effect_2014} and particle-laden \citep{wang_domain_2020} channel flows. Furthermore $Co$ is kept low enough so that aggregates remain smaller than the half channel height in any single spatial dimension to avoid creating infinite chains of particles across the periodic directions and self-interacting wakes. We remark that, given the second-order accuracy of the IBM coupling in the particle-laden cases, the use of a higher-order solver does not yield meaningful benefits.

\subsection{Non-cohesive suspension}
Although no data exists that exactly matches the present configuration, we ensure that the non-cohesive case matches deviations from single-phase flow reported in parametrically nearby cases. \citet{picano_turbulent_2015} investigated turbulence modulation in suspensions of spherical particles of with diameter $2h/16$, density ratio $\rho_s = 1$, at volume fractions $\phi = \{0.05, 0.1, 0.2\}$. For $\phi=0.05$ they report a reduction in streamwise fluctuations and increase in wall-normal and spanwise fluctuations, as also seen here in figure \ref{fig:sp_val}. They also report a similar increase in $\mathit{Re}_\tau$, which they attribute to the formation of a layer of particles at the wall that maintain increased slip velocities due to their finite size. This counteracts the expected decrease in $\mathit{Re}_\tau$ due to reduced turbulence activity. The measured $\mathit{Re}_\tau = 190.05$ for the $Co = 0$ case can also be compared to the scaling law of \citet{costa_scaling_2016}, which estimates a friction Reynolds number taking into account the particle layer at the wall. The error between the scaling law, $\mathit{Re}_{\tau,e}$ and the present DNS is, $1 - \mathit{Re}_\tau / \mathit{Re}_{\tau, e} = 0.037$, which falls within the $4\%$ confidence interval suggested in their work.

\section{The population balance equation}\label{sec:pbe}
An aggregate of size $n$ is defined as a cluster of $n$ particles which are linked through a chain of neighbors separated by a surface distance less than $\lambda$, and thus subject to cohesive forces. The number of aggregates of size $n$, at location $\boldsymbol{x}$, at time $t$ per unit volume is given by the aggregate number density,
\begin{equation}\label{eq:agg-num-dens-point}
    f(n, \boldsymbol{x}, t) = \sum_{i=1}^{N_a(t)}\delta\left(\boldsymbol{x} - \boldsymbol{x}_i^A(t)\right)\delta\left(n - n_i(t)\right),
\end{equation}
where $\boldsymbol{x}_i^A(t) = (x_i^A, y_i^A, z_i^A)$ and $n_i$ are the center of mass and number of primary particles of aggregate $i$, $\delta(\,\cdot\,)$ is the Dirac delta function, and $N_a(t)$ is the total number of aggregates in the domain at time $t$. We take $n$ as the phase-space variable since it is conserved along with the aggregate mass $M_i = n_i \rho_f V_p$, where $V_p = \pi d^3 / 6$ is the volume of a single primary particle (assuming a monodisperse primary particle distribution). Since $n$ is a dimensionless integer, $f$ carries units of $(\text{length})^{-3}$.

Note that $f$ should not be evaluated point wise, but rather only after integrating over a finite region, where it returns the aggregate count. Taking this integration over the full domain $\Omega$ and summing over all $n$ therefore gives,
\begin{equation}
    N_a(t) = \int_\Omega\sum_n f(n, \boldsymbol{x}, t) \,\text{d}V.
\end{equation}
It is not until averaging that \eqref{eq:agg-num-dens-point} yields a smooth, finite probability density.

The evolution of $f$ is described by the PBE,
\begin{equation}\label{eq:pbe-3D}
\frac{\partial nf}{\partial t}
+ \underbrace{\nabla\cdot\left(n\boldsymbol{u}_nf\right)}_{\substack{\boldsymbol{x}\text{-advection}}}
= \underbrace{A_+ + A_-}_{\substack{\text{aggregation}}}
+ \underbrace{B_+ + B_-}_{\substack{\text{breakup}}},
\end{equation}
where $\boldsymbol{u}_n(n, \boldsymbol{x}, t) = (u_n, v_n, w_n)$ is the mean velocity of aggregates of size $n$ in physical space and the $n$ weighting allows for explicit mass conservation. The right side collects source and sink terms $S_i\in\{A_+, A_-, B_+, B_-\}$ that capture discrete aggregation and breakup events, which are assumed binary (two aggregating into one or one fragmenting into two). Nonbinary events are rare and discussed in appendix \ref{ap:fam-tree}. For brevity, explicit dependence on $(\boldsymbol{x}, t)$ is omitted hereafter, and all PBE quantities are understood to carry it unless stated otherwise, e.g. $S_i(n) = S_i(n, \boldsymbol{x}, t)$.

The aggregation source $A_+$ accounts for the formation of size-$n$ aggregates by the coalescence of smaller aggregates,
\begin{equation}\label{eq:A_+}
    A_+(n) = \frac{n}{2} \sum_{m=1}^{n-1} f(m)\, f(n-m)\, \beta(m, n-m)
\end{equation}
where $\beta(m, n-m)$ is the aggregation frequency between two aggregates and the $1/2$ rectifies double-counting. In some PBE formulations $\beta$ can be further partitioned into a product of the collision rate $\alpha(m, n - m)$ and aggregation efficiency $q_a(m, n - m)$, $\beta = \alpha q_a$; however, since all collisions are resolved in this work, a minimum aggregate lifetime would have to be selected to  measure $q_a$, so it is instead assumed to be unity, $q_a = 1$.

The complementary aggregation sink $A_-$ removes size-$n$ aggregates lost to coalescence with any size-$m$ partner,
\begin{equation}\label{eq:A_-}
    A_-(n) = -n\,f(n)\sum_{m=1}^\infty f(m)\, \beta(n, m),
\end{equation}
where the sum spans all sizes since any aggregation event involving the size-$n$ class depletes it. The breakup source term $B_+$ gives the production of size-$n$ fragments from the breakup of larger parent aggregates,
\begin{equation}\label{eq:S_b}
    B_+(n) = n\sum_{m = n+1}^\infty f(m) g_b(m) q_b(m, n),
\end{equation}
where $g_b(m)$ is the breakup rate of size-$m$ parent aggregates and $q_b(m, n)$ is the fragment size distribution. Finally, the sink $B_-$ captures the loss of size-$n$ aggregates due to breakup,
\begin{equation}\label{eq:S_s}
    B_-(n) = -nf(n) g_b(n).
\end{equation}

\subsection{Plane-and-time averaged PBE}\label{sec:plane-pbe}
For channel flow, $f$ is statistically homogeneous in the periodic streamwise and spanwise directions, reducing the spatial dependence to the wall-normal coordinate $y$ alone. Time averaging over a statistically stationary period then yields a PBE dependent only on $(n,y)$,
\begin{equation}\label{eq:pbe-y}
\frac{\partial}{\partial y}\left[\overline{nv_nf}\right](n, y)
= \overline{A_+}(n, y) + \overline{A_-}(n, y) + \overline{B_+}(n, y) + \overline{B_-}(n, y),
\end{equation}
where, as before, a single bar, $\overline{\,\cdot\,}$, denotes a plane-and-time averaged quantity. Crucially, the advective term permits the $\overline{S_i}$ to be locally imbalanced in $y$, allowing for spatially inhomogeneous net aggregation and breakup during equilibrium.

\subsection{Volume-and-time averaged PBE}\label{sec:vol-pbe}
Integrating \eqref{eq:pbe-y} over the channel height and noting that aggregates can not travel through the boundaries eliminates the advective term, leaving a bulk-averaged PBE,
\begin{equation}\label{eq:pbe-global}
    0=\doverline{A_+}(n) + \doverline{A_-}(n) + \doverline{B_+}(n) + \doverline{B_-}(n),
\end{equation}
where a double-bar $\doverline{\,\cdot\,}$ denotes a volume-and-time averaged quantity. What remains is the statement that aggregation and breakup must balance over the entire domain at every $n$.

\subsection{Finite sized particles}\label{subsec:finite-size}
In this work, particles and aggregates are finite-sized, so the point-wise evaluation of equation \eqref{eq:pbe-y} is not possible. Instead, the solid phase volume is partitioned in the wall-normal direction using a grid with spacing $\Delta y = d / 2$ and cell centers $y_j = \Delta y (j + 1 / 2)$  for $j \in \{0, 1, \ldots, 2h / \Delta y - 1 = 61\}$. For a particle $i$ centered at wall-normal coordinate $y_i^P(t)$, the portion of its volume lying within subdomain $j$ is obtained by integrating the spherical volume between the two bounding planes $y_j \pm \Delta y/ 2$. In the frame centered on the particle,
\begin{equation}\label{eq:vp-int}
    \mathcal{V}_i(y_j, t) = \int_{y_{l,i}}^{y_{u,i}}\pi\left(\frac{d^2}{4} - y^2\right)\text{d}y = \pi\left(\frac{d^2}{4}(y_{u,i} - y_{l,i}) - \frac{y_{u,i}^3 - y_{l,i}^3}{3}\right),
\end{equation}
where
\begin{align}
    y_{u,i} &= \min(y_j - y_i^P(t) + \Delta y / 2,\, \phantom{-}d / 2),\\
    y_{l,i} &= \max(y_j - y_i^P(t) - \Delta y / 2,\, -d / 2)
\end{align}
are the subdomain bounds in the wall-normal direction expressed relative to the particle center. The total particle volume in bin $j$ contributed by aggregates of size $n$ is then
\begin{equation}
    V(n, y_j, t) = \sum_{i \,\in\, \mathcal{P}_n(t)} \mathcal{V}_i(y_j, t),
\end{equation}
where $\mathcal{P}_n(t)$ is the set of primary particles belonging to aggregates of size $n$. The discrete, plane-averaged number density follows directly as
\begin{equation}\label{eq:f-discrete}
    \widetilde{f}(n, y_j, t) = \frac{V(n, y_j, t)}{V_p\,V_y},
\end{equation}
where $V_y = L_x L_z \Delta y$ and $\widetilde{\,\cdot\,}$ signifies plane-averaging. Averaging \eqref{eq:f-discrete} over the sampling window yields $\overline{f}(n,y_j)$, the discrete analog of the number density in \eqref{eq:pbe-y}.

Aggregation and breakup events are distributed across the wall-normal direction using the same weighting. The primary particles comprising the parent aggregate (for breakup) or the child aggregate (for aggregation) are located at the moment of the event and their volume is spread across the wall-normal bins using equation \eqref{eq:vp-int}. This ensures $\overline{A_\pm}(n,y)$ and $\overline{B_\pm}(n,y)$ carry the same finite-size treatment as $\overline{f}(n,y)$.

The volume-averaged number density, $\doverline{f}(n, t)$, requires no distribution step, since it does not have spatial dependence and only requires counting the number of size-$n$ aggregates,
\begin{equation}
    \doverline{f}(n, t) = \frac{1}{L_xL_yL_z}\sum_{i=1}^{N_a(t)}\delta(n-n_i(t)).
\end{equation}

\section{Results}\label{sec:results}
\begin{figure}
    \centering
    \includegraphics[trim={0 0.3cm 0 0},clip,width=\textwidth]{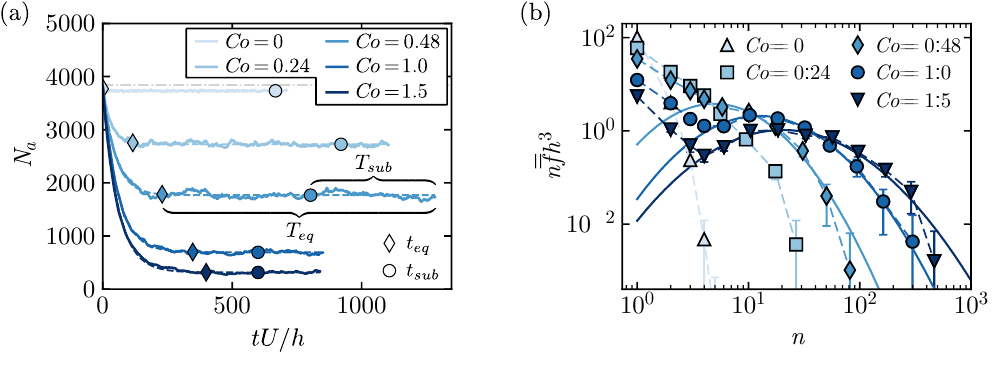}
    \caption{(a) Total number of aggregates $N_a$ over time $t$ for each $Co$. The exponential fit, equation \eqref{eq:N_a_fit}, is shown as dashed lines. The light gray dash-dotted line shows the total number of primary particles $N$. The time to reach equilibrium $t_{eq}$ is marked by diamonds while the start of the subsampling period $t_{sub}$ is marked using circles. The equilibrium and subsampling periods, $T_{eq}$ and $T_{sub}$ respectively are also annotated for case $Co = 0.48$. (b) Mass-weighted aggregate number density $\protect\doverline{nf}$ as a function of size $n$ for each cohesive strength $Co$ (markers with dashed lines). Error bars denote $95\%$ confidence intervals as defined in appendix \ref{sec:error}. Log-normal fits for $n > 3$ are shown as solid lines for $Co \geq 0.48$.}
    \label{fig:Na-fn}
\end{figure}
Simulations first advance through a period of net aggregate growth before reaching a statistically stationary state in which aggregation and breakup are globally balanced. This relaxation toward equilibrium is reflected in the evolution of the total number of aggregates in the domain, $N_a(t)$, shown in figure \ref{fig:Na-fn}(a), adapted from \citet{leonelli_2026}, where a truncated time series of the data considered here is presented. Similar to previously studied cases of sub-Kolmogorov scale particles suspended in HIT \citep{zhao_flocculation_2021} and channel flow \citep{almohammed_modeling_2016}, the relaxation is well-captured by the exponential fit,
\begin{equation}\label{eq:N_a_fit}
    N_a(t) = (N - N_{eq}) e^{-t / k} + N_{eq},
\end{equation}
where the equilibrium number of aggregates, $N_{eq}$, and integral-scale flocculation rate $k$ are fitting parameters. The time to reach equilibrium is then defined as the time it takes $N_a(t)$ to decay to within $1\%$ of $N_{eq}$,
\begin{equation}\label{eq:t_eq}
    t_{eq} = -k\ln\left(\frac{0.01}{N / N_{eq} - 1}\right)
\end{equation}
with the equilibrium period defined as $T_{eq} = [t_{eq}, t_f]$, where $t_f$ is the final snapshot, as illustrated in figure \ref{fig:Na-fn}(a).

During $T_{eq}$, flow and particle statistics are stored every $\Delta t_{IO,eq} = 1.2h/U$. Though appropriately spaced to sample fluid and particle flow statistics, these snapshots are unable to resolve aggregation and breakup events. To do so, particle statistics are stored at a higher rate of $\Delta t_{IO,sub} = 0.01h/U$ during a sub-sampling period $T_{sub} = [t_{sub}, t_f]$ which is also annotated in figure \ref{fig:Na-fn}(a). Within $T_{sub}$, an aggregate family tree is constructed following the rules set in appendix \ref{ap:fam-tree}. Aggregation and breakup events are directly determined from the family tree, with error rates due to polyadic or nonbinary events below 1\% as reported in table~\ref{tab:fam-tree-table}. The total sampling duration is chosen so that uncertainties are sufficiently small as measured using the procedures detailed in appendix \ref{sec:error}.

The bulk mass-weighted aggregate number density, $\doverline{nf}$, is computed using 18 bins in $n$, with four linearly spaced and the remainder logarithmically spaced to balance the discrete nature of small aggregates with sparse statistics at large $n$. Counts in each bin are normalized by the bin width so that $\doverline{nf}$ remains a density in $n$. Results are shown in figure \ref{fig:Na-fn}(b) where, unsurprisingly, larger aggregates are observed for larger $Co$. For $Co = \{0.24, 0.48\}$, $\doverline{nf}$ decreases monotonically, whereas for $Co = \{1.0, 1.5\}$ it increases over an intermediate range of sizes before decaying.

Reported functional forms for the aggregate size distribution vary with flow, material properties, and geometry. Power laws are commonly found in cohesion-dominated systems in HIT \citep{yu_persistent_2023}, with decaying exponentials also commonly reported in both HIT and startup channel flow \citep{zhao_flocculation_2021, almohammed_modeling_2016}.
Experimental and field observations also report a spectrum of size distributions including log-normal \citep{winterwerp_book_2004}. Here we find that a log-normal fit reproduces the non-monotonic behavior at higher $Co$ for $n > 3$ but fails for the smallest aggregates, which remain the most populous class at every $Co$. We therefore treat the fit as descriptive rather than mechanistic, and leave questions on how these distributions emerge to future work. In what follows we instead focus on how the PBE balance varies in the wall-normal direction, where a size-dependent circulation is found to support spatially separated regions of net aggregate production and depletion.

\subsection{Flow modulation}\label{sec:flow}
\begin{figure}
    \centering
    \includegraphics[width=\linewidth]{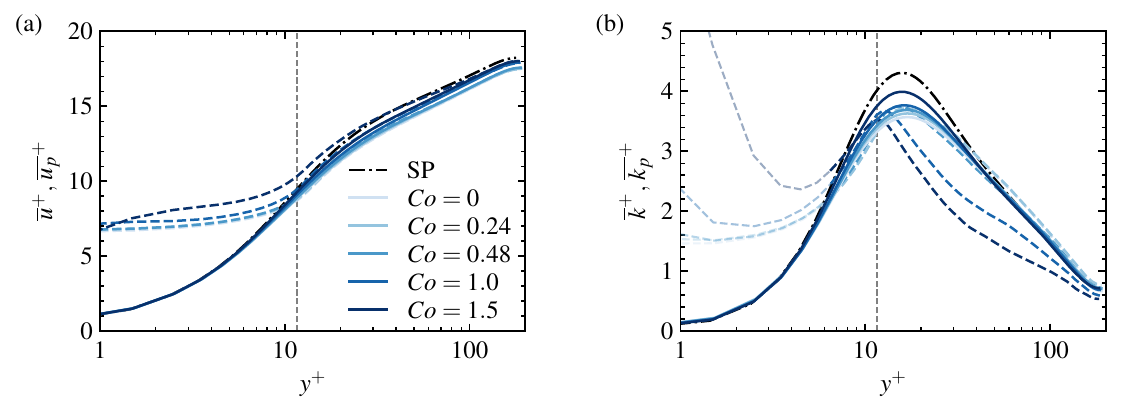}
    \caption{(a) Fluid $\overline{u}^+$ (solid) and particle $\overline{u_p}^+$ (dashed) mean streamwise velocity profiles in inner units for each $Co$. (b) Fluid $\overline{k}^+$ (solid) and particle $\overline{k_p}^+$ (dashed) turbulent kinetic energy profiles in inner units for all $Co$. The single-phase case is included in each panel as a black dash-dotted line while a vertical grey dashed line denotes a particle diameter $d^+$. As discussed in section \ref{sec:flow}, profiles for $\overline{k_p}^+$ are shown with reduced opacity for $y^+ < d^+/2$, where the near-wall singularity in volume fraction makes the plotted (higher-order) statistic increasingly sensitive to sampling noise.}
    \label{fig:vel-prof}
\end{figure}
The mean streamwise fluid and particle velocities, $\overline{u}^+$ and $\overline{u_p}^+$, are plotted as a functions of the wall-normal distance in figure \ref{fig:vel-prof}(a), where the fictitious fluid within particles is removed from the computation of fluid statistics, and the methodology detailed in section \ref{subsec:finite-size} is used to evaluate particle statistics over $N_y = L_y / \Delta x$ wall-normal subdomains. Due to the low mass loading, fluid statistics remain close to the single-phase case (SP) and vary little across $Co$. Within one particle diameter of the wall, $y^+ < d^+ = 11.6$, particles maintain higher streamwise velocities than the fluid owing to their finite size \citep{costa_scaling_2016}, this behavior is likewise insensitive to $Co$.

The turbulent kinetic energy (TKE) for both phases, $\overline{k}^+$ and $\overline{k_p}^+$, are reported in figure \ref{fig:vel-prof}(b). The fluid phase TKE is most prominently suppressed in the buffer layer, a well reported effect in similar particle-laden flows \citep{peng_direct_2019} which appears to reduce with increasing $Co$. For the particle phase TKE, the curves are given a reduced opacity for $y^+ < d^+ / 2$ where, because the local volume fraction must approach zero in proximity of the wall statistics conditioned on the solid phase become increasingly sensitive to sampling noise. This is especially apparent in second or higher order statistics and can not be readily ameliorated due to the singularity in $\phi$ at the wall. Beyond this region, $\overline{k_p}^+$ is generally observed to monotonically increase with $Co$, with $\overline{k_p}^+ < \overline{k}^+$ for larger $Co = \{1.0, 1.5\}$ and $\overline{k_p}^+ \approx \overline{k}^+$ for lower $Co$; an observation consistent with the creation of larger, more inertial aggregates. Importantly, the presence of fluctuations in the particle phase enhances collision rates and aggregate formation \citep{afkhami_effect_2019}, and underpins the aggregation and breakup results presented next.

\subsection{Bulk PBE balance}\label{sec:bulk}
When considering the full domain volume, the PBE is balanced solely by aggregation and breakup, as expressed by equation \eqref{eq:pbe-global}.
\begin{figure}[ht]
    \centering
    \includegraphics[width=\linewidth]{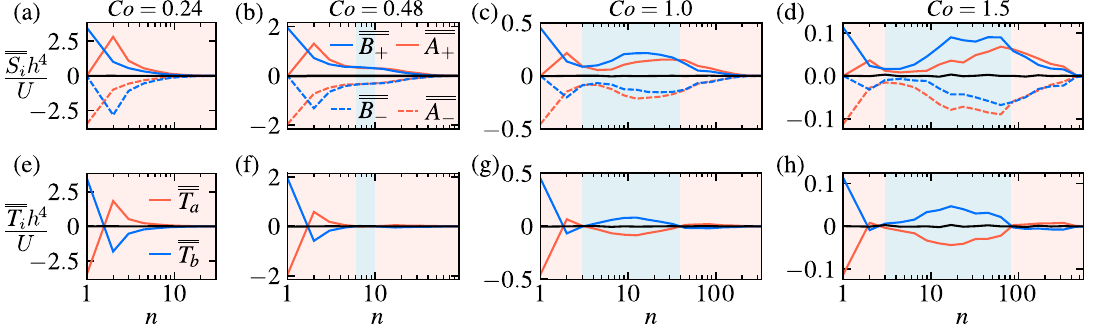}
    \caption{Volume- and time-averaged aggregation and breakup source terms, $\protect\doverline{S_i}$ (a-d) and net aggregation and breakup source terms $\protect\doverline{T_a}$ and $\protect\doverline{T_b}$ (e-h). Each column shows data for a single, nonzero cohesive strength. The $\protect\doverline{S_i}$ are defined in section \ref{sec:pbe} while $\protect\doverline{T_a}$ and $\protect\doverline{T_b}$ are given by equations \eqref{eq:T_a} and \eqref{eq:T_b} respectively. Sizes $n > 1$ where $\protect\doverline{T_a} > \protect\doverline{T_b}$ are highlighted with a red background whereas those with $\protect\doverline{T_b} > \protect\doverline{T_a}$ have blue backgrounds.}
    \label{fig:ST_bulk}
\end{figure}
The measured bulk source terms, $\doverline{S_i}(n)$, for the cases $Co > 0$ are shown in Figure \ref{fig:ST_bulk} (a--d) across size-space. Trivially, $\doverline{A_+} = \doverline{B_-} = 0$ at $n=1$ for all $Co$, since monomers cannot be produced by aggregation nor break. It follows that at $n=1$, $\doverline{B_+} = -\doverline{A_-}$, reflecting that monomers are produced solely by breakup of larger aggregates and are removed only through aggregation.

At larger sizes, $\doverline{A_+}$ and $\doverline{B_-}$ become active as mass is transferred bidirectionally in size-space. The sum of all contributions, $\doverline{Q} = \sum \doverline{S_i}$, shown in black, is approximately zero for all cases, consistent with \eqref{eq:pbe-global}, and demonstrating conservation of mass across size-space.

To further aid interpretation, the net aggregation and breakup fluxes,
\begin{equation}\label{eq:T_a}
    \doverline{T_a} = \doverline{A_+} + \doverline{A_-},
\end{equation}
\begin{equation}\label{eq:T_b}
    \doverline{T_b} = \doverline{B_+} + \doverline{B_-},
\end{equation}
are shown as red and blue curves in panels (e--h), respectively. The quantity $\doverline{T_a}$ represents the net gain or loss of size $n$ aggregates due to aggregation alone. $\doverline{T_a}(n) > 0$ indicates that formation from smaller sizes exceeds loss to larger sizes, implying a net aggregation-driven flux of mass into $n$ from below. Conversely, $\doverline{T_a} < 0$ implies a net transfer of mass out of $n$ towards larger aggregates. $\doverline{T_b}$ is analogously constructed for breakup. $\doverline{T_b} > 0$ indicates a net breakup-driven flux of mass into $n$ from above while $\doverline{T_b} < 0$ indicates a net transfer of mass from $n$ to smaller sizes.

Taken together, $\doverline{T_a}$ and $\doverline{T_b}$ provide a compact representation of the local directional mass fluxes in size-space, isolating the competing roles of aggregation-driven mass flux to larger $n$ and breakup-driven flux downscale. As before, the combined rate $\doverline{Q} = \doverline{T_a} + \doverline{T_b} = \sum \doverline{S_i}$ is shown as a black line. Deviations from 0 should be interpreted as error since particle mass is conserved over the full domain.

With this statistical formulation in mind, we revisit panels (e-f) in figure \ref{fig:ST_bulk}. At the lowest nonzero cohesion, $Co=0.24$ (panels (a,e)), $\doverline{T_a} \geq 0 \geq \doverline{T_b}$ for all $n > 1$ with $\doverline{T_a} = -\doverline{T_b}$ as required by \eqref{eq:pbe-global}. Thus, aggregation acts to bring mass to all $n>1$ while breakup acts as a sink to balance. Restated, for all $n > 1$ aggregation of smaller aggregates into size $n$ outcompetes further aggregation of size $n$ into larger sizes, thus $\doverline{T_a} > 0$. Necessarily, $\doverline{T_b} \leq 0$ indicating that breakup of size $n$ to smaller scales outpaces incoming fragments of size $n$ from the breakup of larger aggregates. In contrast, the strongest cohesion, $Co = 1.5$ (panels (d,h)), displays a significant range of sizes where $\doverline{T_b} \geq 0 \geq \doverline{T_a}$ indicating that these $n$ receive mass from larger sizes ($\doverline{T_b} \geq 0$) and, on-average, redistribute it upscale $\doverline{T_a} \leq 0$. This region is highlighted using a blue background for each panel. Inspecting panel (d) it is apparent that $\doverline{B_+}$ is balanced by $\doverline{A_-}$ in this region, revealing a statistical cycle that limits further fragmentation. Finally, a complementary range of sizes where $\doverline{T_a} \geq 0 \geq \doverline{T_b}$ is necessarily present at larger $n$ and is highlighted by a red background.

The range of $n$ where $\doverline{T_b} \geq 0 \geq \doverline{T_a}$ grows smaller with decreasing $Co$ and is not present at $Co=0.24$, where cohesion is too weak to interrupt a statistically continuous transfer of mass downscale due to breakup (i.e. breakup is a sink for all $n > 1$). The exact relationship between these regions and $Co$ is complex and likely flow and geometry dependent.

At this point we have verified that the simulations have reached a statistically stationary state for which the global population balance satisfies $\doverline{Q}(n) = 0$ for all $n$, consistent with  \eqref{eq:pbe-global}. However, it is unknown if this is the case locally in the wall-normal direction.

\subsection{Local wall-normal balance}\label{sec:wall-normal-balance}
When the PBE is plane-averaged, as in \eqref{eq:pbe-y}, the advective term allows for transport of aggregates in the wall-normal direction, hence aggregation and breakup are not required to locally balance in $y$. To assess this, we employ the methodology described in section \ref{subsec:finite-size} to partition the channel into 62 wall-normal subdomains of thickness $\Delta y = d/2$ and evaluate the net source term,
\begin{equation}\label{eq:Q}
    \overline{Q}(n, y) = \overline{T_a}(n, y) + \overline{T_b}(n, y),
\end{equation}
for three size classes, $n=1$, $n=2,3$, and $3 < n < 20$. These subsamples must be considered individually since aggregation and breakup conservatively redistribute mass across sizes (i.e. a breakup event at $n=3$ will be balanced by the breakup source term at $n=1$ and $n=2$ in the same $y$ bin), such that
\begin{equation}
\sum_{n=1}^\infty \overline{Q}(n, y) = \sum_{n=1}^\infty \overline{T_a}(n, y) = \sum_{n=1}^\infty \overline{T_b}(n, y) = 0.
\end{equation}
Similarly, in the wall-normal direction,
\begin{equation}
    \int_0^{L_y}\, \overline{Q}(n, y)\text{d}y = 0,
\end{equation}
which is a restatement of equation \eqref{eq:pbe-global}.

\begin{figure}[ht]
    \centering
    \includegraphics[trim={0 0.2cm 0 0.2cm},clip,width=\linewidth]{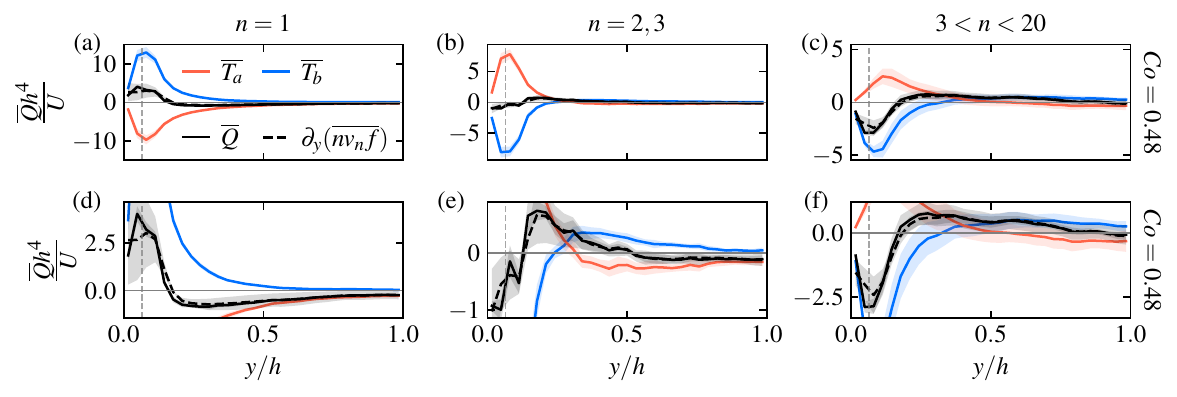}
    \caption{The net source due to aggregation and breakup, $\overline{Q}$, aggregation flux $\overline{T_a}$, breakup flux $\overline{T_b}$, and advective flux $\partial_y(\overline{nv_nf})$, for three size classes: $n = 1$, $n \in \{2,3\}$, and $3 < n < 20$ (columns) for $Co = 0.48$ plotted over the half channel height, $y/h \in [0, 1]$. The upper row (a-c) shows the full data range while the bottom row focuses on $\overline{Q}$ and $\partial_y(\overline{nv_nf})$(d-f). A 95\% confidence interval is shown by the shaded region surrounding each curve. The gray vertical dashed line corresponds to $y=d$.}
    \label{fig:Q_Co0p48}
\end{figure}

The resulting net source term, $\overline{Q}(n, y)$, is shown over half the channel, $y/h\in[0, 1]$ (exploiting symmetry about $y=h$) in figure \ref{fig:Q_Co0p48} for $Co = 0.48$. $\overline{Q}(n, y)$ is clearly nonzero, demonstrating that aggregation and breakup are not locally balanced in the wall-normal direction. The top row of figure \ref{fig:Q_Co0p48} shows the full range of data, while the bottom row places focus on $\overline{Q}$ to highlight imbalance between $\overline{T_a}$ and $\overline{T_b}$. Each curve is accompanied by a shaded region showing a 95\% confidence interval computed as described in appendix \ref{sec:error}.

For monomers ($n=1$, panels (a,d)), aggregation and breakup trivially act as a sink and source respectively, thus $\overline{T_a} \leq 0$ and $\overline{T_b} \geq 0$ for all $y$. Near the wall breakup supplies mass to the $n=1$ class faster than aggregation removes it, thus, $\overline{Q} > 0$. Away from the wall, this trend is reversed ($\overline{Q} < 0$) as the removal due to aggregation dominates.

For larger sizes ($3 < n < 20$) the dynamics are equally straightforward as $\overline{T_a} \geq 0$ and $\overline{T_b} \leq 0$ for all $y$. The near wall region of $\overline{Q} < 0$ is therefore a consequence of breakup-driven depletion outpacing aggregation-driven supply whereas away from the wall breakup slows due to the reduced turbulence intensity and stresses while aggregation is less affected, so that there is a net mass influx to these sizes ($\overline{Q} > 0$) near the channel center.

The dynamics are more intricate for intermediate sizes ($n=2, 3$, panels (b, e)). Near the wall, aggregation acts as a source ($\overline{T_a} > 0$), breakup as a sink ($\overline{T_b} < 0$), and the net balance is negative ($\overline{Q} < 0$), indicating that breakup-driven loss exceeds aggregation-driven gain. Moving away from the wall, $\overline{Q}$ becomes positive, implying that aggregation now dominates the balance. Interestingly, within this region, aggregation and breakup as source and sink invert as $y$ increases. Toward the channel center, $\overline{Q} < 0$ again, but here depletion is driven by a net mass flux toward larger sizes as aggregation outpaces the local supply from breakup.

These observed imbalances in aggregation and breakup must be sustained by the wall-normal advection of aggregates as stated in \eqref{eq:pbe-y}. Indeed, the resultant advective flux, shown in figure \ref{fig:Q_Co0p48} (black dashed line), is in good agreement with $\overline{Q}$, demonstrating closure of the PBE in the wall-normal direction.

\begin{figure}
    \centering
    \includegraphics[trim={0 0.2cm 0 0.2cm},clip,width=\linewidth]{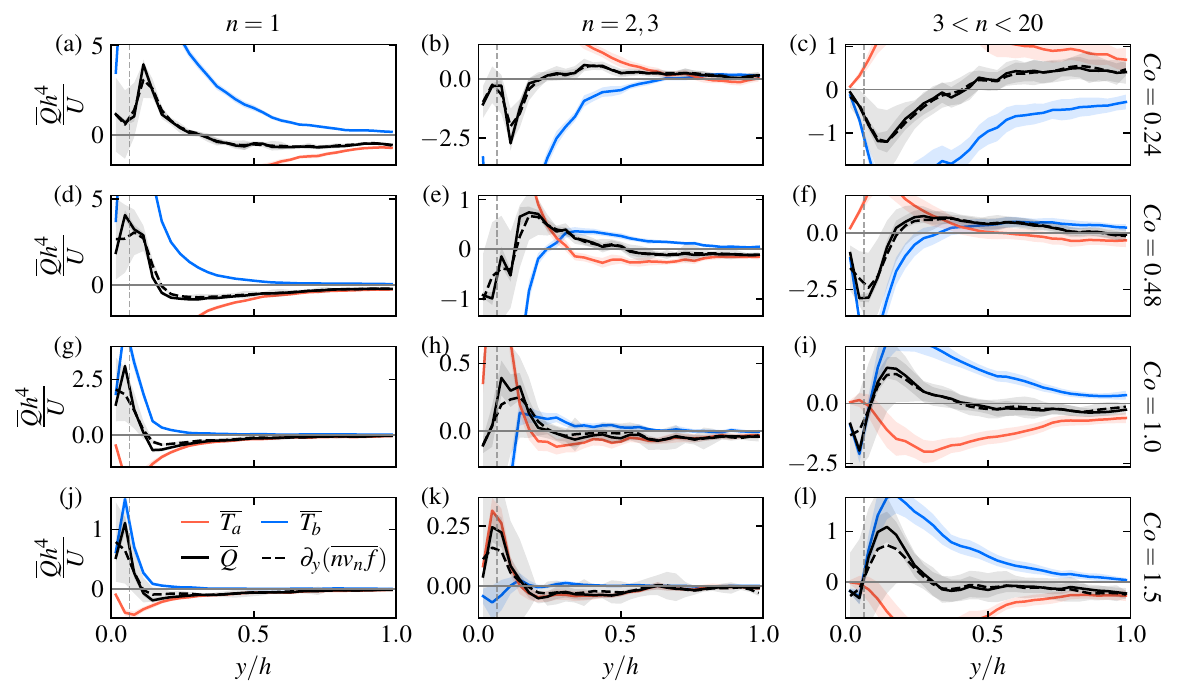}
    \caption{The net source due to aggregation and breakup, $\overline{Q}$, aggregation $\overline{T_a}$, breakup $\overline{T_b}$, and advective flux $\partial_y(\overline{nv_nf})$ for three size classes: $n = 1$, $n \in \{2,3\}$, and $3 < n < 20$ (columns) for $Co > 0$ (rows) plotted over the half channel height, $y/h \in [0, 1]$. As in figure \ref{fig:Q_Co0p48} (d-f) the axes limits are chosen to focus $\overline{Q}$; a full figure is given in appendix \ref{sec:all-source}. A 95\% confidence interval is shown by the shaded region surrounding each curve. The gray vertical dashed line corresponds to $y=d$.}
    \label{fig:all-source-zoomed}
\end{figure}

\begin{figure}
    \centering
    \includegraphics[width=\linewidth]{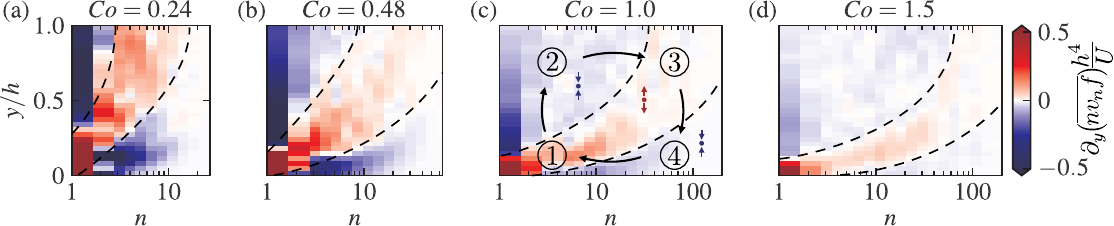}
    \caption{Wall-normal advective transport of aggregates, $\partial_y(\overline{nv_nf})$ in $(n, y)$ space for each $Co > 0$ (a-d). Dashed lines are drawn onto the plot to aid in distinguishing between regions of $\partial_y(\overline{nv_nf}) > 0$ (red) and $\partial_y(\overline{nv_nf}) < 0$  (blue). Panel (c) has additional annotation that is used in the text to describe the cyclic transfer of mass through regions 1-4. Symbols with vertical arrows show the divergence of $\overline{nv_nf}$ to compliment the red and blue coloring.}
    \label{fig:dvf_dy_heatmap}
\end{figure}

Results for all $Co > 0$ are presented in figure \ref{fig:all-source-zoomed}. Just as for $Co = 0.48$, $\overline{Q}(n, y) \neq 0$, showing that aggregation and breakup are locally imbalanced for all $Co > 0$. The same general trends for $n=1$, $n=2,3$ and $3 < 20$ are present across $Co$. Near the wall, monomers are created in excess due to breakup while they are depleted away from the wall by aggregation. For $Co = \{0.24, 0.48\}$, large aggregates ($3 < n < 20$) are on average sourced in the channel center by aggregation and break near the wall. For higher $Co = \{1.0, 1.5\}$, this observation is reversed as aggregates sized $3 < n < 20$ are seen be sourced near the wall by breakup and depleted in the channel center by aggregation, behaving similarly to the $n=1$ class for lower values of $Co$. For $Co = \{1.0, 1.5\}$ the complimentary range of sizes in which aggregation acts as a source in the channel center resides at much larger $n$ as shown later in figure \ref{fig:dvf_dy_heatmap}. Intermediate sizes, exhibit net production or depletion that depends on $Co$: at lower $Co$, they behave similarly to the $3 < n < 20$ class, whereas at higher $Co$, they resemble monomers, as they are unable to break in the channel center while continuing to be depleted by aggregation. As for $Co = 0.48$, the advective term is necessary to maintain this inhomogeneity in $\overline{Q}$ and close the PBE. Notably, \citet{peng_direct_2019} observed that non-cohesive particles may form aggregate-like clusters dependent on the lubrication model employed. While varying parameters and models is beyond the scope of this work it should be noted given the relative importance of the near-wall region and lubrication interactions demonstrated here.

Having identified regions of net production and depletion, and aware of the intricate interplay between aggregation, breakup, and advection, we now examine how these dynamics organize in $(n, y)$ space. Plotting the wall-normal advective flux (or $\overline{Q}$) in $(n, y)$ space reveals a statistical circulation of aggregates across size and physical space, illustrated in figure \ref{fig:dvf_dy_heatmap} with four annotated regions in (c). Regions in which the advective flux serves to remove aggregates are red whereas those which accumulate mass are shown in blue. Beginning at region 1, near the wall at small $n$, the advective flux removes mass. Since no aggregates can be transported through the wall at $y=0$, this net outflow must carry mass away from the wall toward the channel center, into region 2. There, the advective flux accumulates mass. This region 2 spans from its boundary with region 1 to the channel center. Because region 2 begins at $n=1$, the advectively accumulated mass cannot be redistributed to smaller sizes and therefore must be transferred upward in size-space into region 3. Region 3 exhibits an advective outflow and spans the channel center, so mass is advected from it toward the wall, into region 4. Note that downscale transfer out of region 3 is statistically precluded, since region 2 already accumulates mass by advection. Region 4 as well accumulates mass by advection, but unlike region 2 it spans from its lower boundary with region 1 up to the maximum aggregate size. Consequently, the inflow from the channel center (region 3) must be balanced by a downscale redistribution of mass back into region 1. Together, these four regions define a closed statistical cycle: small aggregates near the wall are advected toward the channel center, where they grow by aggregation into larger sizes, before migrating back toward the wall to break up and repeat the cycle.

This cyclical structure in $(n, y)$-space appears at all $Co$ in figure \ref{fig:dvf_dy_heatmap}(a-d), with each region extending to larger $n$ as $Co$ increases. The $y$-span of region 1 decreases with increasing $Co$, reflecting that aggregates of a given size are more resilient and thus to break, must travel closer to the wall where stresses are highest. A final subtlety is that the dependence of these trajectories on $Co$ indicates that they are governed by aggregation and breakup, rather than morphology-dependent transport. In other words, the net production or depletion of aggregates of a given size within a region is controlled by population gradients and not necessarily the propensity of a particular size to migrate toward the channel center or wall.

\subsection{Wall interactions and aggregate fate}\label{sec:agg-fate}
The circulation identified in figure~\ref{fig:dvf_dy_heatmap} is anchored by region-1, where near-wall breakup depletes the local population of large aggregates and seeds the advective outflow. We now ask what sets the fate of an aggregate that reaches this region, and in particular whether direct contact with the wall, the lubrication model (equation \eqref{eq:lub-model}), or general hydrodynamic stresses drive breakup near the wall.

A particle-wall collision occurs for particle-wall separations $\zeta_{iw} \leq 0$, with small negative values possible due to the soft-sphere collision model. The lubrication model, which acts in the wall-normal direction only, is active for $0 < \zeta_{iw} \leq 2\Delta x$ while outside this range, the hydrodynamics are fully resolved. To separate these, we define for each aggregate $j$ the smallest particle--wall gap among its set of constituent particles $\mathcal{A}_j$,
\begin{equation}\label{eq:zeta-agg}
    \zeta^A_j(t) = \min_{i\,\in\,\mathcal{A}_j} \zeta_{iw}(t).
\end{equation}
The lifespan of each aggregate terminates by either an aggregation or breakup event at its time of death $t_{d,j}$. For every aggregate, we find the closest approach to the wall before $t_{d,j}$ over a window $\tau_d$,
\begin{equation}\label{eq:zeta-event}
    \zeta_{min,j} = \min_{\tau \in [0,\,\tau_d]} \zeta^A_j(t_{d,j} - \tau).
\end{equation}
Events are then assigned to one of the three mutually exclusive modes according to $\zeta_{min,j}$: ``collisional'' ($\zeta_{min,j} \leq 0$), ``lubricated'' ($0 < \zeta_{min,j} \leq 2\Delta x$), and ``hydrodynamic'' ($2\Delta x < \zeta_{min,j} \leq 4\Delta x$). Here we choose the ``hydrodynamic'' region to be of equal thickness to the lubrication layer so that the aggregate populations within each region are relatively homogeneous. We use $\tau_d = h/U$ throughout, noting that results remain relatively insensitive to this choice provided $\tau_d > 0.5h/U$ (see appendix \ref{app:window}).

\begin{figure}
    \centering
    \includegraphics[width=0.5\linewidth]{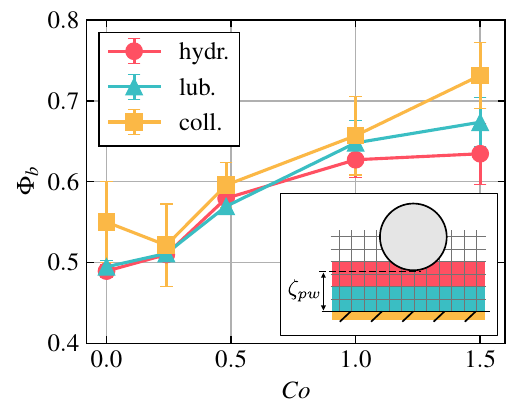}
    \caption{Proportion of aggregates that break over the total number of aggregation and breakup events assigned to each mechanism $\Phi_b$ plotted over cohesive strength $Co$; particle-wall collisions, particle-wall lubrication, and shear-induced breakup. Error bars denote a 95\% confidence interval. A schematic showing the definition of the regions relative to the grid spacing is shown as an inset where $\zeta_{pw}$ is the distance between the surfaces of particle and wall. Note that the particle is drawn smaller relative to the grid spacing than in reality.}
    \label{fig:break-region}
\end{figure}
Within each region we report the breakup fraction $\Phi_b = N_{break} / (N_{break} + N_{agg})$, which gives the proportion of aggregates that break over the total number of events classified in that region. Figure~\ref{fig:break-region} shows $\Phi_b$ for each breakup classification across all $Co$. This analysis applied to the $Co = 0$ case yields $\Phi_b \approx 0.5$ for all breakup categories. This is unsurprising considering that without a cohesive force, inter-particle collisions mostly begin (aggregation) and terminate (breakup) in the same region so that $N_{agg} \approx N_{break}$ locally in the wall-normal direction. With increasing $Co$ the ratio of breakup to aggregation events increases, consistent with the conceptual picture established in section~\ref{sec:wall-normal-balance} that breakup out-competes aggregation in the near-wall region. Notably, $\Phi_b$ increases monotonically with $Co$ independent of any region-specific mechanism (collisional, lubricated, hydrodynamic). This follows from the aggregate size distribution itself (figure \ref{fig:Na-fn}(b)) where for increasing $Co$ a larger fraction of the population resides in large aggregates, depleting the pool of small aggregates available as aggregation partners. At intermediate cohesive strengths, $Co = \{0.24, 0.48, 1.0\}$, $\Phi_b$ appears roughly equal for all breakup categories, suggesting that no single interaction disproportionally drives breakup. Only for the most cohesive case, $Co=1.5$, do the categories separate nearly beyond the $95\%$ confidence interval denoted by the error bars, with $\Phi_b\approx 0.63$ for general hydrodynamic breakup, $\Phi_b\approx 0.67$ for lubrication, and $\Phi_b\approx 0.73$ for particle-wall collisions. This modest but ordered separation ($\Phi_b^{coll} > \Phi_b^{lub} > \Phi_b^{hydr}$), indicates that only at the strongest cohesive strength, $Co=1.5$, aggregates that interact directly with the wall by either the normal lubrication model or direct contact are increasingly likely to break rather than aggregate. This effect is minimal, especially considering that hydrodynamic forces are still acting on the aggregates in the lubrication and particle-wall collision interactions. Taken together, these results suggest that the fate of an aggregate that enters close enough proximity to the wall to interact by the modeled normal lubrication force is relatively unaffected when compared to those that do not, suggesting that breakup is generally facilitated by hydrodynamic stresses.

\subsection{Aggregate orientation and axial straining}\label{sec:axial_strain_rate}
To further understand how aggregate dynamics facilitate breakup, we first examine the wall-normal profiles of aggregate orientation. We then show that aggregates maintain rotation rates similar to the fluid, facilitating their alignment with the streamwise- ($x$) axis. Finally, we show that the rate of strain along the aggregate long axis increases near the wall, demonstrating that aggregates become progressively more extended on approach to the wall, promoting breakup.

\begin{figure}
    \centering
    \includegraphics[width=\linewidth]{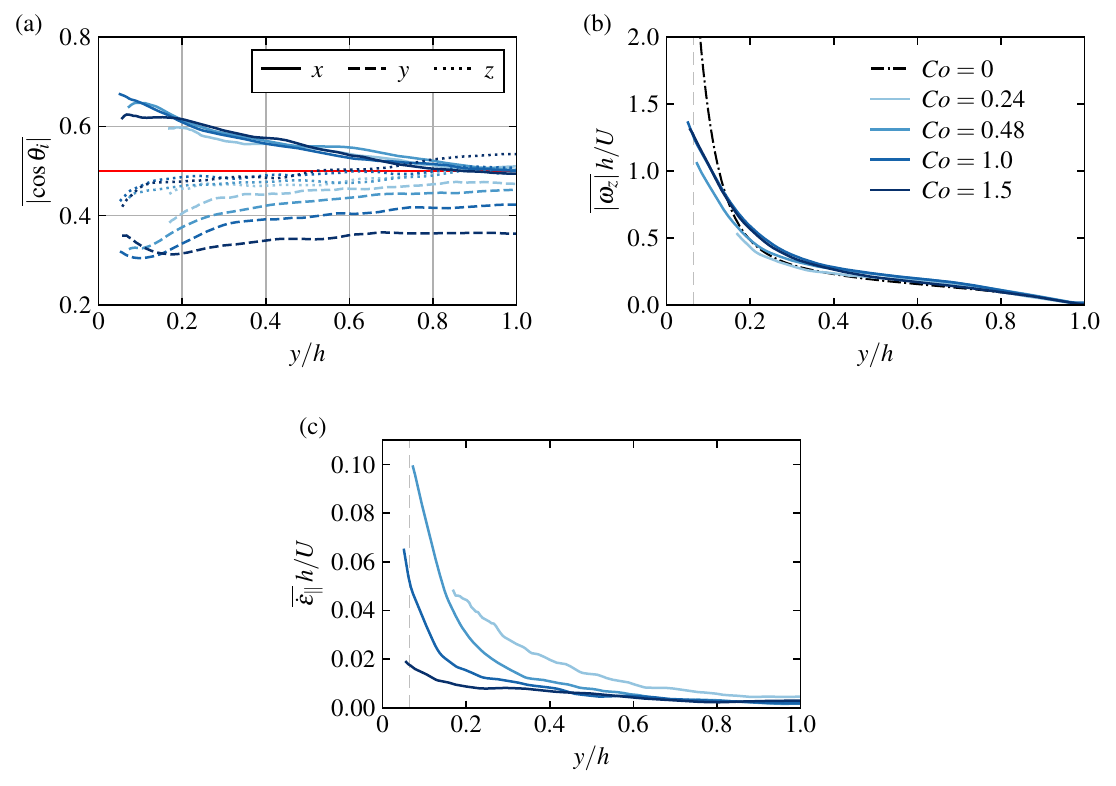}
    \caption{Wall-normal profiles of: (a) Orientation angle of the aggregate longest axis with respect to the domain frame of reference, equation \ref{eq:theta}, (b) the angular velocity of aggregates about the spanwise axis, $z$, (c) mean rate of strain along the aggregate long axis. Each figure only considers aggregates with $n>2$ and wall-normal bins with $\phi > 10^{-3}$ to avoid singularities in particle phase conditioned statistics.}
    \label{fig:strain}
\end{figure}

We define the orientation of an aggregate $j$ to be the unit principal eigenvector of its gyration tensor, $\boldsymbol{e}^A_j$, as defined in appendix \ref{app:agg_shape}. The alignment of aggregate $j$ with the domain coordinate triad, $\boldsymbol{e}_i$ with $i\in\{x, y, z\}$ is then,
\begin{equation}\label{eq:theta}
    \cos\theta_{i,j} = \lvert \boldsymbol{e}^A_j \cdot \boldsymbol{e}_i \rvert.
\end{equation}
By construction \eqref{eq:theta} is agnostic to the sign of $\boldsymbol{e}_j^A$, since it is arbitrary. It follows that $\lvert\cos\theta_i\rvert \rightarrow 1$ indicates alignment with axis $i$, while $\lvert\cos\theta_i\rvert \rightarrow 0$ indicates misalignment. Averaging over all aggregates size $n > 2$ sampled during $T_{eq}$ and following the binning procedures in section \ref{subsec:finite-size} for $N_y$ subdomains gives the mean orientation, $\overline{\cos\theta_i}$. For an isotropically oriented population, $\overline{\cos\theta_i} = 0.5$ in every direction $i$. The wall-normal profiles of the orientation for aggregates with $n > 2$ are shown in figure \ref{fig:strain}(a), where small aggregates are omitted since $\boldsymbol{e}_j^A$ becomes ill-defined. One issue that arises when excluding small aggregates is that the population becomes sparse for low-$Co$ cases, yielding noisy statistics due singularities that arise when the local volume fraction is small. To avoid this, only subdomains where $\overline{\phi}(y_j) > 10^{-3}$ are considered, which is why the data for $Co = 0.24$ is truncated.

In the channel center, aggregate alignment appears generally isotropic, $\lvert \cos\theta_i\rvert\approx0.5$, with slight misalignment along the $y$-axis emerging at higher $Co$. Approaching the wall, aggregates align preferentially with the streamwise direction, with $\lvert\cos\theta_x\rvert\rightarrow0.7$ for all $Co$, accompanied by a corresponding suppression of wall-normal alignment. Spanwise alignment remains isotropic ($\cos\theta_z \approx 0.5$) throughout.

Since aggregate alignment varies with wall-normal position, and aggregates are simultaneously transported across $y$, an aggregate carried through this gradient must itself rotate about the spanwise axis to track the local mean orientation, with angular velocity $\omega_z$. As shown in figure \ref{fig:strain}(b), the aggregate rotation rate tracks the fluid rotation rate exceptionally well, with slip apparent only in close proximity to the wall, $y\lesssim 0.15h$. No significant mean rotation is observed about the wall-normal or streamwise axes, which are not reported here since $\overline{\lvert\omega_x\rvert}\approx\overline{\lvert\omega_y\rvert}\approx 0$.

Given that aggregates rotate and become preferentially aligned with the streamwise direction near the wall, an appropriate deformation measure is the rate of strain along $\boldsymbol{e}_j^A$. Computed following the methodology in appendix \ref{app:agg_shape}, the axial rate of strain, $\dot\varepsilon_\parallel$, is averaged in the same manner as the orientation and reported in figure \ref{fig:strain}(c). A positive $\dot\varepsilon_\parallel$ indicates axial extension, while negative values denote compression.

The mean profiles $\overline{\dot\varepsilon_\parallel}(y)$ are shown in figure~\ref{fig:strain}(c) for each $Co$. Aggregates are extended near the wall at every $Co$, with the extensional straining increasing monotonically on approach to the wall, where breakup is most prominent. The apparent decrease of $\overline{\dot\varepsilon_\parallel}$ very close to the wall at $Co = 0.24$ reflects the scarcity of aggregates of size $n > 2$ there, rather than a change in mechanism. Satisfactorily, the magnitude of $\overline{\dot\varepsilon_\parallel}$ decreases monotonically with $Co$ at fixed $y$ and reflects that stronger cohesion makes aggregates stiffer so that a given shear rate yields less axial stretching. This is consistent with the observations of near-wall breakup mechanisms in figure \ref{fig:break-region} where collision and lubrication forces were found to increasingly contribute to breakup relative to the shear at high $Co$.

Together, these mechanistic observations corroborate the statistical interpretation: breakup is facilitated in the near wall region, where a mixture of hydrodynamic, lubrication, and collisional forces facilitate aggregate deformation and breakup. More cohesive aggregates (higher $Co$) are deformed less at a given wall-normal location and therefore must travel closer to the wall before breaking, consistent with the contraction of the $y$-span of region~1 with increasing $Co$ observed in figure~\ref{fig:dvf_dy_heatmap}.

This analysis is a first interrogation of the mechanisms sustaining the newly framed statistical circulation in $(n, y)$ space. A predictive account requires the full wall-normal and size dependence of the PBE kernels themselves, together with their scaling with local flow quantities; a substantial task that is left to future work.

\section{Conclusion}\label{sec:conclusion}
In this work we have brought forward five PR-DNS of a turbulent channel flow laden with finite sized cohesive particles, with each case characterized by increased cohesive strength, $Co$. Initially randomly dispersed, the particles form aggregates, break, and redistribute, until a statistically stationary population is reached, with higher $Co$ yielding larger aggregates and non-monotonic mass-weighted number densities.

Due to the low volume fraction and neutrally buoyant particles flow modulation is shown to be minimal. Despite this, the fluctuations of the particle phase are graded by $Co$, reflecting the growth of large inertial aggregates. Two-way coupling between aggregates and the carrier flow deserves closer attention than was given here, particularly in light of the preferential streamwise alignment of aggregates near the wall reported here; an effect which has been shown to have dramatic consequences on turbulent statistics, including wall shear stress, in other anisotropic particles such as fibers and spheroids \citep{voth_anisotropic_2017, marchioli_particle-laden_2025}. Though not significant here, such effects should become important for heavier particles, or for smaller primary particles, where aggregates may deform substantially before breaking.

Unlike sub-Kolmogorov scale particles or aggregates, the finite-size nature of the structures considered in this work will likely present size-dependent kinematics due to spatial filtering, where larger aggregates will effectively filter out relatively smaller scales and kinematically respond only to similar or larger scale flow structures \citep{coletti_filtering_2025}. As a result the flocculation rate may be lower due to the reduction in fluctuations, as reported in figure \ref{fig:vel-prof}. Unfortunately, the statistics measured in this work are insufficient to further investigate this, and direct evidence for size-dependent filtering is left as a subject of future work.

Near the wall, aggregates are found to align with the flow, with rotation rates closely tracking those of the fluid. They are also shown to undergo increasingly rapid extension along their long axis on approach to the wall, where breakup is prominent. Classifying breakup events by closest approach to the wall shows that, at low $Co$, direct contact and lubrication modeling do not alter the outcome (aggregation or breakup) of a near-wall encounter relative to hydrodynamic forces alone, whereas for the strongest cohesion, $Co = 1.5$, they do, consistent with the reduced deformability of strongly cohesive aggregates. The breakup of larger aggregates in this near-wall region results in an excess supply of small aggregates that must be removed if equilibrium is to be maintained, a fate that can be tracked statistically with the PBE.

Exploiting the PBE analytically, we show that aggregation and breakup must balance globally, since together they conservatively redistribute mass across sizes. Nontrivially, however, they need not balance locally in the wall-normal direction. Equilibrium is instead maintained through spatially separated regions of net aggregate production and depletion and closure of the PBE is recovered with the inclusion of size-dependent wall-normal advective transport. Viewed in $(n, y)$-space, these regions organize into a closed statistical cycle: small aggregates are produced in abundance near the wall, advected towards the channel center where they grow through aggregation, transported back toward the wall, and subsequently broken down, re-entering the cycle.

\paragraph*{Acknowledgements.}
A.D.L. and E.M. thank Dr. Ronald Chan and Prof. Filippo Coletti for many insightful discussions. A.D.L. and E.M. are grateful for their participation in the 2024 Stanford Center for Turbulence Research Summer Program at which this work was initiated. Financial support was received from the Geological Survey of Israel, the Army Corps of Engineers through grant No. W912HZ22C0037, and the Army Research Office through grant No. W911NF-23-2-0046. This work used Anvil at Purdue University RCAC through allocation MCH260012 and CTS150053 from the Advanced Cyberinfrastructure Coordination Ecosystem: Services \& Support (ACCESS) program, which is supported by U.S. National Science Foundation grants \#2138259, \#2138286, \#2138307, \#2137603, and \#2138296.

\paragraph*{Declaration of interests.}
The authors declare no conflicts of interests.

\paragraph*{Data availability statement.}
The data that support the findings of this study are available upon request.

\appendix
\section{Slip velocity}\label{sec:slip-vel}
\begin{figure}
    \centering
    \includegraphics[width=0.45\textwidth]{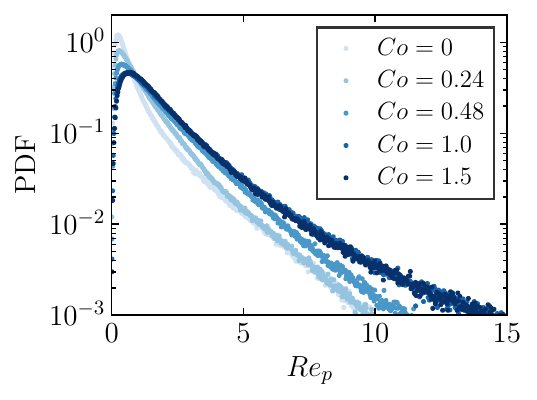}
    \caption{Probability density distributions of particle Reynolds number $\mathit{Re}_p$ measured over all aggregates in the sampling period $T_{eq}$ for each $Co$.}
    \label{fig:Re_p}
\end{figure}

The slip velocity between particle and fluid is computed by taking the mean fluid velocity within a shell of radius $R = d / 2 + 3\Delta x$ centered at particle $p$,
\begin{equation}
    \boldsymbol{u}_i^F(t) = \frac{\displaystyle\int_0^R \int_0^{2\pi}\int_0^\pi \boldsymbol{u}(\boldsymbol{x}_i^P(t) + \boldsymbol{r}, t)\,\xi_f(\boldsymbol{x}_i^P(t) + \boldsymbol{r}, t)\, r^2\sin\theta\,\text{d}r\, \text{d}\theta\, \text{d}\phi}{\displaystyle\int_0^R \int_0^{2\pi}\int_0^\pi \xi_f(\boldsymbol{x}_p(t) + \boldsymbol{r}, t)\, r^2\sin\theta\,\text{d}r\, \text{d}\theta\, \text{d}\phi},
\end{equation}
where $\boldsymbol{r} = \boldsymbol{r}(r,\theta,\phi)$ denotes the position vector, in spherical coordinates, relative to $\boldsymbol{x}_i^P$. Here, the fluid indicator function $\xi_f$,
\begin{equation}
    \xi_f(\boldsymbol{x}, t) = \begin{cases}
        0 & \text{if inside a particle},\\
        1 & \text{otherwise},
    \end{cases}
\end{equation}
ensures that no fluid from inside the fictitious domain within particles is included in the estimate for $\boldsymbol{u}^F_i$. The slip velocity for a particle $i$ is then given by $\Delta\boldsymbol{u}^{PF}_i = \boldsymbol{u}^P_i - \boldsymbol{u}^F_i$ which is then used to estimate the particle Reynolds number, $\mathit{Re}_{p,i} = \Delta\boldsymbol{u}^{PF}_i d / \nu$. In figure \ref{fig:Re_p}, the PDF of $\mathit{Re}_p$ accumulates $\mathit{Re}_{p,i}$ for each case averaged over all particles and $T_{eq}$. Expectedly, the highest $\mathit{Re}_p$ are observed for the largest $Co$ due to the creation of larger aggregates. The maximum $\mathit{Re}_{p,i}$ over all cases occurs for $Co = 1.5$ and is estimated as $\max_i\mathit{Re}_{p,i} = 49.58$, which is the value used in section \ref{sec:config} to estimate the minimum particle boundary layer thickness.

\section{PBE computation}\label{ap:pbe-compute}
\subsection{Family tree}\label{ap:fam-tree}
Aggregates are independently identified at every stored snapshot. A family tree is then generated by matching the constituent primary particles between snapshots. Four outcomes are possible when constructing continuity between $t$ and elapsed time $t + \Delta t_{IO,sub}$:
\begin{itemize}
    \item \textbf{No change}: the same set of primary particles remain as one component at $t + \Delta t_{IO,sub}$.
    \item \textbf{Aggregation event}: an aggregate is identified at time $t + \Delta t_{IO,sub}$ that is comprised of two (binary) or more (non-binary) aggregates from the current time $t$.
    \item \textbf{Breakup event}: the primary particles, and only the primary particles, that comprised a single aggregate at time $t$ now constitute two (binary) or more (non-binary) aggregates identified at time $t + \Delta t_{IO,sub}$.
    \item \textbf{Polyadic event}: the primary particles of two or more aggregates at time $t$ are intermingled and comprise two or more aggregates at time $t + \Delta t_{IO,sub}$.
\end{itemize}

Due to the ability to fully resolve the constituent particle kinematics, we only consider binary aggregation and breakup events in this work, and consider all other events as ``missed'' binary events. The identification rate of each type of event is reported in Table~\ref{tab:fam-tree-table} where it is shown that less than $1\%$ of events are missed over all cases.

\begin{table}
    \centering
    \begin{tabular}{lllllll}
        Case &
        \shortstack{Resolved\\agg.} &
        \shortstack{Resolved\\break.} &
        \shortstack{Missed\\agg.} &
        \shortstack{Missed\\break.} &
        \shortstack{\% missed\\agg.} &
        \shortstack{\% missed\\break.} \\
        \hline
        $Co = 0$ & 3431  & 3426  & 7   & 11  & 0.20 & 0.32 \\
        $Co = 0.24$ & 14713 & 14727 & 68  & 82  & 0.46 & 0.56 \\
        $Co=0.48$ & 26767 & 26641 & 161 & 247 & 0.60 & 0.93 \\
        $Co=1.0$ & 4746  & 4707  & 21  & 30  & 0.44 & 0.64 \\
        $Co=1.5$ & 1610  & 1621  & 9   & 11  & 0.56 & 0.68 \\
    \end{tabular}
    \caption{Aggregation and breakup event detections. Missed events are either nonbinary or polyadic as defined in section \ref{ap:fam-tree}.}
    \label{tab:fam-tree-table}
\end{table}

\section{Correlation time and error estimation}\label{sec:error}
Statistical uncertainty is reported in the main text using a $95\%$ confidence interval. To compute these intervals the number of independent snapshots, $N_{eff}$, taken over the sampling period for each $Co$ case is first estimated, then used to compute the standard error.

The statistical interpretation of turbulence is strictly based on ensemble averages, where each snapshot is independent. This is impractical in many turbulent flows as lead time can be costly. Instead, Taylor's hypothesis is invoked and sample independence is justified using the correlation times of fluid properties such as the streamwise velocity fluctuations. In this work, we must additionally quantify timescales related to the rearrangement of aggregate populations so that the number of independent samples of the aggregate number density and related quantities can be estimated. To do so, we evaluate the autocorrelation of the fluctuations of the global mass-weighted aggregate number density $\dwidetilde{nf}(n, t)$, where $\dwidetilde{\,\cdot\,}$ denotes a mean taken over the full computational domain $\Omega$. The fluctuations are then,
\begin{equation}
    \dwidetilde{nf}'(n, t) = \dwidetilde{nf}(n, t) - \doverline{nf}(n),
\end{equation}
where the time and volume averaged number density $\doverline{nf}$ is recovered from $\dwidetilde{nf}$ by time averaging.

\subsection{Global autocorrelation and correlation time}
\begin{figure}
    \centering
    \includegraphics[width=0.5\textwidth]{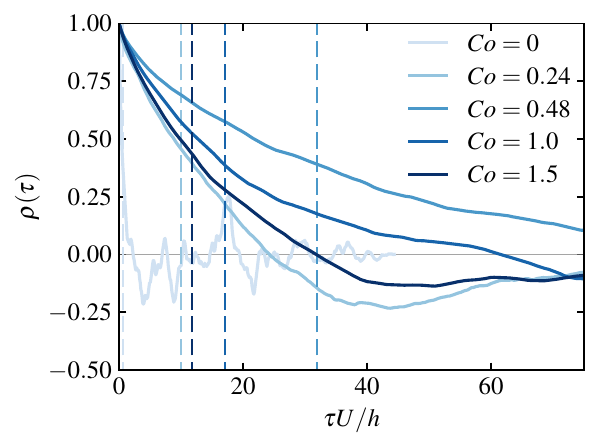}
    \caption{Global autocorrelation as defined by equation \eqref{eq:corr} for all $Co$. The correlation time, $\tau_{int}$, is plotted as a vertical dashed line of matching color for each case.}
    \label{fig:corr}
\end{figure}

We are interested in the fluctuations of the distribution as a whole and therefore define the global autocorrelation as,
\begin{equation}\label{eq:corr}
    \rho(\tau) = \frac{\sum_n\left\langle \dwidetilde{nf}'(n, t) \dwidetilde{nf}'(n, t + \tau)\right\rangle}{\sum_n \left\langle \dwidetilde{nf}'(n, t)\dwidetilde{nf}'(n, t)\right\rangle},
\end{equation}
where $\langle\,\cdot\,\rangle$ denotes a time average. Due to their abundance, this formulation is dominated by the statistics for $n=1$ aggregates. However, the $n=1$ population is generally observed to have the longest correlation times and so we take this as a suitable conservative estimate.
The resulting $\rho(\tau)$ are plotted for every case in figure \ref{fig:corr}. The correlation time,
\begin{equation}\label{eq:tau}
    \tau_{int} = \int_0^{\tau_0} \rho(\tau) \text{d}\tau,
\end{equation}
quantifies the rate of relaxation of $\rho(\tau)$ and is shown as a vertical dashed line for each case. Here, $\tau_0$ is the first zero crossing of $\rho(\tau)$, which is taken as the upper limit of integration to avoid integration of random noise at large $\tau$.

\subsection{Number of samples and error estimation}
The number of independent samples is estimated by dividing the total sampling period for each case $T$ by $2\tau_{int}$,
\begin{equation}
    N_{eff} = \frac{T}{2\tau_{int}},
\end{equation}
where the factor of 2 follows from the standard relationship between the integral autocorrelation time and the effective number of statistically independent samples in a stationary process. The number of independent samples is reported in table \ref{tab:results} alongside the correlation times and sampling periods.

To estimate the uncertainty on a measurement such as the aggregate number density $\doverline{nf}(n)$ we quantify the variance $\text{Var}[nf(n, t)]$ and compute a standard error,
\begin{equation}\label{eq:se}
    SE = c_{95}\sqrt{\frac{\text{Var}[nf(n, t)]}{N_{eff}}},
\end{equation}
where $c_{95}$ is the prefactor for a 95\% confidence interval. Given by the two-tailed critical value of the Student's $t$-distribution with $N_{eff}-1$ degrees of freedom, $c_{95} = t_{0.975, \, N_{eff}-1}$. This choice accounts for the fact that $N_{eff}$ is small for several cases (as low as 8), where the Gaussian approximation $c_{95} \approx 1.96$ underestimates the true interval width.
\begin{table}
    \centering
    \begin{tabular}{llrrrrrr}
    \toprule
    $Co$ & $T_{sample}$ & $T_{sub}$ & $\tau_{int}$ & $N_{eff, samp}$ & $N_{eff, sub}$ & $c_{95, sample}$ & $c_{95, sub}$ \\
    \midrule
    0    & 708.43  & 44.37  & 0.57  & 618 & 39 & 1.96 & 2.02 \\
    0.24 & 987.83  & 183.78 & 9.94  & 50  & 10 & 2.01 & 2.26 \\
    0.48 & 1053.34 & 480.68 & 31.88 & 17  & 8  & 2.12 & 2.36 \\
    1.0  & 501.94  & 249.60 & 17.14 & 15  & 8  & 2.14 & 2.36 \\
    1.5  & 439.40  & 238.82 & 11.69 & 19  & 11 & 2.10 & 2.23 \\
    \bottomrule
    \end{tabular}
    \caption{Timing statistics results and error prefactors.}
    \label{tab:results}
\end{table}

\section{PBE fluxes, expanded}\label{sec:all-source}
\begin{figure}[ht]
    \centering
    \includegraphics[trim={0 0.2cm 0 0.2cm},clip,width=\linewidth]{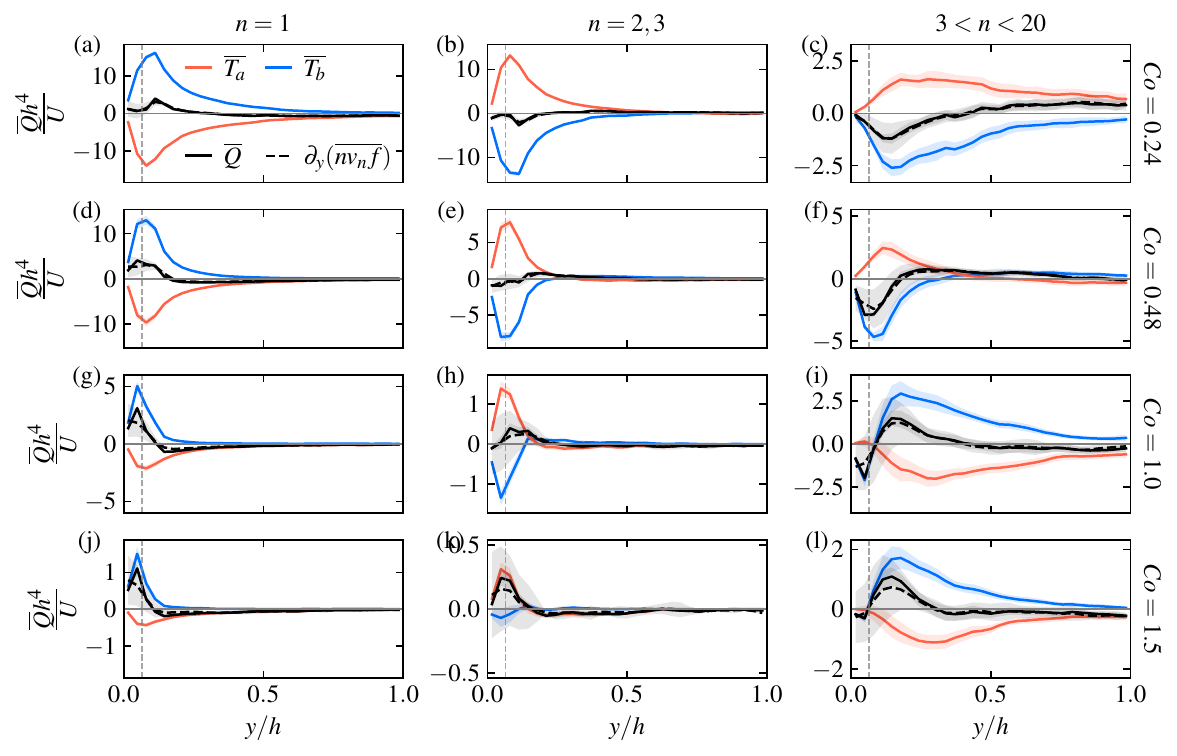}
    \caption{The net source due to aggregation and breakup, $\overline{Q}$, aggregation $\overline{T_a}$, breakup $\overline{T_b}$, and advective flux $\partial_y(\overline{nv_nf})$ for three size classes: $n = 1$, $n \in \{2,3\}$, and $3 < n < 20$ (columns) for $Co > 0$ (rows) plotted over the half channel height, $y/h \in [0, 1]$. A 95\% confidence interval is shown by the shaded region surrounding each curve. The gray vertical dashed line corresponds to $y=d$.}
    \label{fig:all-source}
\end{figure}

Figure \ref{fig:all-source-zoomed} in section \ref{sec:wall-normal-balance} presents the wall-normal profile of the net source due to aggregation and breakup $\overline{Q}(y)$ with the data for the individual net source and sink terms due to aggregation and breakup $\overline{T_a}$ and $\overline{T_b}$ cut from view due to the section's focus on $\overline{Q}$. Here, we show the full range of data in figure \ref{fig:all-source}.

\section{Sensitivity to window}\label{app:window}
\begin{figure}
    \centering
    \includegraphics[width=\linewidth]{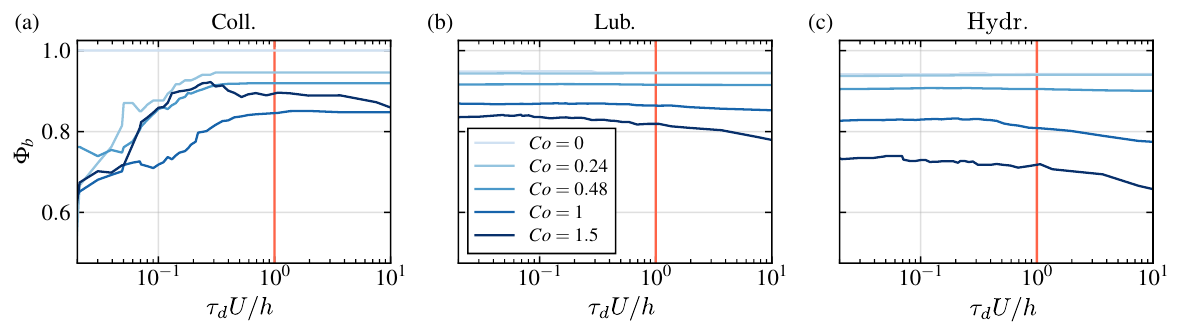}
    \caption{Sensitivity of breakup fraction $\Phi_b$ for each breakup mechanism (a) collisional, (b) lubricated, or (c) hydrodynamic, as a function of cutoff time $\tau_d$. The selected $\tau_d = h / U$ used in section \ref{sec:agg-fate} is shown as a vertical red line.}
    \label{fig:tau_d_sensitivity}
\end{figure}

The cause of breakup in section \ref{sec:agg-fate} is determined based on an aggregate's closest approach to the wall in a window $\tau_d$ before its time of death, $t_d$. In the main text $\tau_d = h/U$. Here we plot the breakup ratio $\Phi_b$ as defined in section \ref{sec:agg-fate} for each mechanism (collisional, lubricated, and hydrodynamic) as a function of $\tau_d$ in figures \ref{fig:tau_d_sensitivity}(a-c). Over all $Co$ it is clear that $\Phi_b$ is relatively insensitive to $\tau_d$ so long as $\tau_d\gtrsim 0.5h/U$. The selected value of $\tau_d = h/U$ is shown as a vertical red line in figures \ref{fig:tau_d_sensitivity}(a-c).

\section{Aggregate shape and orientation}\label{app:agg_shape}
We take the long axis of an aggregate $j$, $\boldsymbol{e}_j^A$ to be the unit principal eigenvector of the gyration tensor,
\begin{equation}
    \mathcal{G}_j = \frac{1}{n_j}\sum_{i=1}^{n_j} \boldsymbol{r}^{PA}_i \left(\boldsymbol{r}^{PA}_i\right)^T,
\end{equation}
where $\boldsymbol{r}^{PA}_i = \boldsymbol{x}^P_i - \boldsymbol{x}^A_j$ is the position of particle $i$ relative to the center of mass of its host aggregate $j$, $\boldsymbol{x}_j^A$. Since the sign of an eigenvector is arbitrary, the orientation is calculated using
\begin{equation}\label{eq:theta-app}
    \cos\theta_{j,k} = \left\lvert \boldsymbol{e}_k \cdot \boldsymbol{e}^A_j \right\rvert,
    \qquad k \in \{x, y, z\},
\end{equation}

To then compute the rate of strain along $\boldsymbol{e}_j^A$, first, the position and velocity of each constituent particle is projected onto the axis, $r_{\parallel,ij} = \boldsymbol{r}_i^{PA}\cdot\boldsymbol{e}_j^A$ and $u_{\parallel,ij} = \boldsymbol{u}_i^P\cdot\boldsymbol{e}_j^A$. We next define the least-squares estimate of the axial velocity gradient across the aggregate,
\begin{equation}
    \dot\varepsilon_{\parallel,j} = \frac{\sum_i u_{\parallel,ij}r_{\parallel,ij}}{\sum_i r_{\parallel, ij}^2}
\end{equation}
for $i\in\mathcal{A}_j$ where $\mathcal{A}_j$ is the subset of particles in aggregate $j$. Here, positive $\dot\varepsilon_\parallel$ indicates axial extension and negative values axial compression.

\bibliographystyle{plainnat}
\bibliography{references}

\end{document}